\documentclass[sn-mathphys-num]{sn-jnl}

\usepackage{graphicx}%
\usepackage{multirow}%
\usepackage{amsmath,amssymb,amsfonts}%
\usepackage{amsthm}%
\usepackage{mathrsfs}%
\usepackage[title]{appendix}%
\usepackage{xcolor}%
\usepackage{textcomp}%
\usepackage{manyfoot}%
\usepackage{booktabs}%
\usepackage{algorithm}%
\usepackage{algorithmicx}%
\usepackage{algpseudocode}%
\usepackage{listings}%

\usepackage{subcaption}
\usepackage{float}
\usepackage{booktabs}
\usepackage{url}
\usepackage{caption}
\usepackage{natbib}
\usepackage{pdflscape}
\usepackage{afterpage}

\theoremstyle{thmstyleone}%
\theoremstyle{thmstyletwo}%

\theoremstyle{thmstylethree}%

\begin{document}


\title{Implications of construction decisions in keyword-based networks: an empirical assessment}

\author*[1]{\fnm{James} \sur{Nevin}}\email{j.g.nevin@uva.nl}

\author[2]{\fnm{Salvatore Flavio} \sur{Pileggi}}\email{salvatoreflavio.pileggi@uts.edu.au}

\author[1]{\fnm{Michael} \sur{Lees}}\email{m.h.lees@uva.nl}

\author[1]{\fnm{Paul} \sur{Groth}}\email{p.t.groth@uva.nl}

\affil*[1]{\orgdiv{Informatics Institute}, \orgname{University of Amsterdam}, \orgaddress{ \city{Amsterdam}, \country{Netherlands}}}

\affil[2]{\orgdiv{Faculty of Engineering and IT}, \orgname{University of Technology Sydney}, \orgaddress{\city{Sydney}, \country{Australia}}}


\abstract{The large amounts of data continuously generated online offer opportunities to identify and analyse trends in various aspects of society. For instance, data from online social media are frequently used as a means of analysing informal interactions, opinions, and feelings of groups of people. Additionally, bibliometric data can be used to investigate more formal trends that occur in scientific research. A popular approach to analysing such complex semi-structured data is the construction of complex networks based on keywords or concept extraction. However, such keyword-based complex network data are often shared in a preprocessed form, with little information about the underlying process used to construct it. Indeed, key decisions are normally made at an early stage in the construction of complex networks from raw data, and can have a significant impact on subsequent analysis and interpretation. In this paper, we highlight the sensitivity of results to data preprocessing decisions by looking at two different case studies which employ networks constructed from underlying semi-structured data. The experiments conducted show high sensitivity to data preprocessing for many commonly adopted metrics. These results demonstrate the need for transparent reporting of data lineage and preprocessing decisions.}




\maketitle

\section{Introduction}\label{sec:int}

Large amounts of data are created through digital communication and interaction. These interactions are diverse and complex, and this complexity is reflected in the datasets generated from traces of these digital interactions. Creating and analysing models from such datasets require systematic approaches that are both interpretable and scalable \citep{bono2023pipeline,pastrav2020norms,fischer2005socionics,kim2016examples}.

One source of these interactions is online social media platforms, which offer opportunities to model and interpret how people think, interact, and make decisions. The use of social media data has become a mainstay in many areas of science, from sociology, to epidemiology, to economics \citep{varshney2021review, gunti2022data, kosinski2014manifestations, pourmand2019social, prasad2019purchase}.

Similarly, research can be done on the scientific literature. While the data generated by online social media platforms is usually fairly unregulated and informal, scientific articles undergo rigorous review prior to being published. At the same time, much like with social media data, such articles also offer insight into the issues being studied and thus what is considered important.

While both types of data offer many opportunities, they are generally large, unstructured text datasets that need to be converted into forms that can be systematically analysed. A typical approach to do this is to create complex networks. These complex networks transform data into interpretable structures \citep{cohen2010complex}. Such transformations are done through data preprocessing pipelines, and these pipelines can dramatically impact the resulting networks. 

A common approach to defining a complex network based on text data is through the use of keywords \citep{garg2018structure,fudolig2022sentiment,lozano2019complex}. Keywords can be extracted from the text and used to define different types of complex network. In word co-occurrence networks, words/terms are nodes and edges connect terms that occur together in the same social media post/scientific article. Another possibility is to define nodes based on users/authors or posts/articles and connect them based on the similarity in terms used. Such networks allow one to identify important posts, which authors act as bridges in conversations, and more.

The construction of such networks does not necessarily follow a standard process, and hence requires significant preprocessing of raw data. The process is sensitive to a number of decisions in the different phases such as, for example, what constitutes a term or how to address filler words. Additionally, different tools may be adopted, and the current generation of AI-powered approaches normally require the setting of parameters. As a consequence, there are multiple potential networks that could be constructed from the exact same raw dataset. The challenge is that it is hard to understand a priori how diverse these potential networks are and which preprocessing decisions have the most influence.

Due to dataset sizes, privacy concerns, or restrictions from data owners, in most cases it may not be possible for researchers to share their raw, unprocessed data; instead, they may share only the fully constructed network. However, as data preprocessing decisions will dictate the exact form of the network, there is the potential for results to look very different depending on the preprocessing applied. Hence, there is a need for transparency, with the exact decisions and methods adopted throughout the data engineering process being disclosed. This further implies the need to consider multiple data preprocessing techniques and whether results are highly sensitive to decisions in order to provide a reliable and well-rounded analysis.

Our previous work in this area \citep{nevin2021non, nevin2023approach, nevin2023data} demonstrated, at a high-level, the sensitivities of complex networks and diffusion models executed on networks to data handling decisions. Additionally, it proposed an approach to systematically characterise these sensitivities. Although the data handling concerns in the mentioned works do not perfectly correspond to the data preprocessing decisions tested in this paper, the same analysis framework can be adopted.

In this paper, we delve deeper into the data preprocessing issue for social media and bibliometric data by creating word co-occurrence networks using different data preprocessing approaches. For social media data, we use a set of tweets by public figures; for bibliometric data, we use the titles and abstracts of articles on climate change from Scopus. We test an automatic keyword extraction technique to create these networks, where multiple setups are considered in terms of length of keyword. We find that global network properties (such as degree centrality distributions), node rankings based on these properties, and community structure show sensitivity to the choice of data preprocessing.

The remainder of this paper is organised as follows: the following section provides some necessary background material about social media and bibliometric analysis and introduces the datasets (Section~\ref{sec:background}), methodological aspects are the object of Section~\ref{sec:methodology}, while results are presented and discussed in Sections~\ref{sec:results} and \ref{sec:discussion}, respectively; finally, the paper ends with conclusions that also address future work. 

\section{Background}\label{sec:background}

This section aims first to provide a concise overview of two key concepts: (i) Complex network analysis, and (ii) Keyword-based complex networks. Following this, we describe the datasets used and their characteristics.

\subsection{Complex network analysis}

Complex network analysis has progressively emerged as a powerful tool for identifying critical patterns and behaviours in large systems through the identification and definition of relatively simple relationships among entities \citep{dorogovtsev2022nature, boschetti2005defining, delgado2002emergence}. 

Complex network analysis has achieved significant maturity \citep{mata2020complex} and many of its consolidated tools have been applied to the more specific fields of social and bibliometric networks.

Complex networks are defined as graphs made up of pairs of nodes and edges \citep{barabasi2013network}. Nodes represent entities and edges connections between these entities. Both nodes and edges can be defined in different ways, leading to different networks and subsequent analysis and insights.

These networks can be analysed according to different methodologies and metrics, including centrality, partitioning, and global properties such as diameter \citep{staudt2016networkit}. In particular, distributions of centrality are often used to characterise complex networks into classes, such as the so-called `scale-free' \citep{barabasi2003scale} and `small-world' \citep{amaral2000classes} networks. These centrality measures can further be used in identifying influential nodes \citep{sciarra2018change, salavati2019ranking}. The definition of specific metrics and their application is an open area of research \citep{wang2022identifying, ugurlu2022comparative}, while traditional measures, such as degree and closeness centrality, are largely used in practice \citep{fudolig2022sentiment, malik2022complex}.

\subsection{Keyword-based complex networks}

Complex networks created using keywords have been applied to both social media data and bibliometric data. These keywords (also known as `n-grams') can be made up of more than one word, such as `climate change'. Keywords are either defined by users/authors (e.g. hashtags in tweets or author keywords in articles) or can be extracted from the text in a post or article.

There are a number of examples of using keywords in analysing social media data. Using word co-occurrence networks, Mocibob \emph{et al.} \cite{mocibob2016revealing} investigated the structure of tweets within a specific domain; Garg and Kumar \cite{garg2018structure} looked into the structure of word co-occurrence networks constructed from microblogs; Fudolig \emph{et al.} \cite{fudolig2022sentiment} considered the network structures arising under different sentiments in co-occurrence networks in political tweets. These prior analyses have generally used user-defined keywords (hashtags). There are some limitations to this approach, such as room for manipulation by users (hijacking of popular hashtags \citep{fudolig2022sentiment}) or non-transferability to other social media sources with different post structure. 

An alternative approach is keyword extraction. Moreno-Ortiz and Garc{\'\i}a-G{\'a}mez \cite{moreno2023strategies} offer a review of various different strategies to keyword extraction from tweets and the strengths and weaknesses of each. Jayasiriwardene and Ganegoda \cite{jayasiriwardene2020keyword} show the use of natural language processing to extract keywords from tweets and collect relevant news. These keyword extraction approaches avoid some of the issues of the user-defined keywords, while introducing other concerns that need to be taken into account.

Keyword-based complex networks have also been used in the study of bibliometric data. Lozano \emph{et al.} \cite{lozano2019complex} searched Web of Science articles for efficiency analysis studies and analysed word co-occurrence networks based on author-defined keywords; Li \emph{et al.} \cite{li2016evolutionary} also used Web of Science articles and author-defined keywords to evaluate the evolution of networks based on keyword co-occurrence and the articles in which they appear. In these and other cases using author-defined keywords, there is still a substantial amount of preprocessing work to do, such as removing stop words, addressing acronyms, etc. Additionally, the quantity of data available is growing, which means that the traditional approach of  manual preprocessing is becoming intractable.

Once again, automatic keyword extraction addresses some of these concerns \citep{radhakrishnan2017novel}. For example, a large-scale study by Fu and Waltman \cite{fu2022large} investigated trends in climate change research. The authors used a natural language processing approach to extract keywords from titles and abstracts.

While there are advantages to automatic keyword extraction in both the social media and bibliometrics domains, there are a number of challenges in construction of networks based on them: how do we extract meaningful keywords within posts/articles? What makes one word more important than another? How do we identify `noisy' keywords? Should they be removed? These are semantic questions that have downstream sensitivity consequences. Given the huge amount of data normally processed and the consequent need for quantitative approaches, consideration also has to be given to scalability and interpretability. All of these types of questions and challenges need to be either implicitly or explicitly addressed by researchers when engineering keyword-based complex networks. We refer to the transformations along the process that translates raw data to complex networks as \emph{data preprocessing}, i.e., decisions that affect the engineering process outcome.

In all of the previous studies described above, only one data preprocessing was applied. This runs the risk of finding results that are not robust.

\subsection{Datasets}\label{sec:data}

Here, we introduce the two datasets and some of the characteristics and challenges associated with each.

\subsubsection{Social media data}

The social media dataset is the \textit{Twitter Parliamentarian Database}, originally introduced by Van Vliet \emph{et al.} \cite{van2020twitter}. It includes tweets from parliamentarians within the member states of the European Free Trade Association, as well as a number of majority English-speaking countries such as the United States and Canada. The published data consists of tweet IDs that can be rehydrated to collect the original tweets.

This dataset has already been used in part in a study relating to polarisation around climate change \citep{falkenberg2022growing}, while similarly defined datasets have been used for studying polarisation and democracy \citep{esteve2022political,praet2021patterns}. Additionally, specific work into word co-occurrence networks has collected tweets on political topics such as war and immigration \citep{mocibob2016revealing}.

A further advantage of this dataset is its high level of openness, being that it is generated by public parties. This limits the privacy and ethical concerns that must be taken into account. Considering all these aforementioned factors, this dataset offers well-rounded possibilities for analysis.

The latest version of the original dataset at the time of writing has a total of 11358559 tweets. After rehydrating the first 10000 tweets, Twitter`s API \footnote{\url{https://developer.twitter.com/en/docs/twitter-api}. Accessed: April 2023.} policy was changed, which made rehydrating tweets prohibitively expensive, which also limited the possibilities of expanding the data set. However, a similar number of tweets were collected for similar studies \citep{mocibob2016revealing}.

Of the original 10000 selected tweets, a total of 8616 were still accessible at the time of rehydration (April 2023). The missing tweets could no longer be available for a number of reasons, including deletion and change in visibility settings. Additionally, a number of tweets were truncated as it was not possible to recover the full text. For each tweet, the raw data collected included a tweet ID, an author ID, and the tweet text.

\subsubsection{Bibliometric data}

The bibliometric data used is similar to that of Fu and Waltman \cite{fu2022large}. Article titles and abstracts are collected from Scopus, which offers comparable results as using Web of Science \citep{fu2022large,torres2009ranking,archambault2009comparing}. The following climate change related search terms were used: `climatic chang*', `climate variabilit*', `climatic variabilit*', `climate warming', `climatic warming', `climate chang*', and `global warming'. Searches were conducted for the period 2001 to 2018, inclusive.

There were a total of 139968 unique articles found from which we extracted the titles and abstracts. As in Fu and Waltman \cite{fu2022large}, we selected a random subset of 25000 articles, which is enough to find robust results.

\subsubsection{Dataset characteristics}

The results of the analysis performed using automated keyword extraction are heavily influenced by the characteristics of the dataset used. 

For the Twitter dataset, there are a number of key properties that have an impact:
\begin{enumerate}
    \item The tweets are multilingual. This affects the sentiment analysis and the likelihood that keywords occur.
    \item The tweets are general and not topic-based. Since longer keywords tend to be more specific, it is less likely that longer keywords will occur together.
    \item The dataset is relatively small, which also limits how well-connected keywords can be.
\end{enumerate}

These properties make it difficult to draw decisive conclusions. With this in mind, our focus is on highlighting the sensitivities of various network measurements to the data preprocessing decisions. Because of this, the analysis we perform is extensive, but not exhaustive. Specifically, we only illustrate a subset of results, noting that one can delve far deeper with a more complete dataset. While results are not generalisable, they do function well as an initial illustrative/theoretical example.

The Scopus data does not have these same limitations. It is a full dataset that is both far more detailed and topic-focused, and it is thus generally possible to find many connections between keywords. However, the larger size does mean that it is not computationally feasible to work with all the data, and so only a subset can be used.

\section{Methodology and approach}\label{sec:methodology}

This section provides an overview of methodological aspects, including a detailed description of the experiments conducted. More concretely, it addresses keyword extraction and sentiment analysis (Section~\ref{sec:data_processing}), network engineering (Section~\ref{sec:engineering}), and network analysis (Section~\ref{sec:analysisMethod}).

\begin{figure}
    \centering
    \includegraphics[width=1\textwidth]{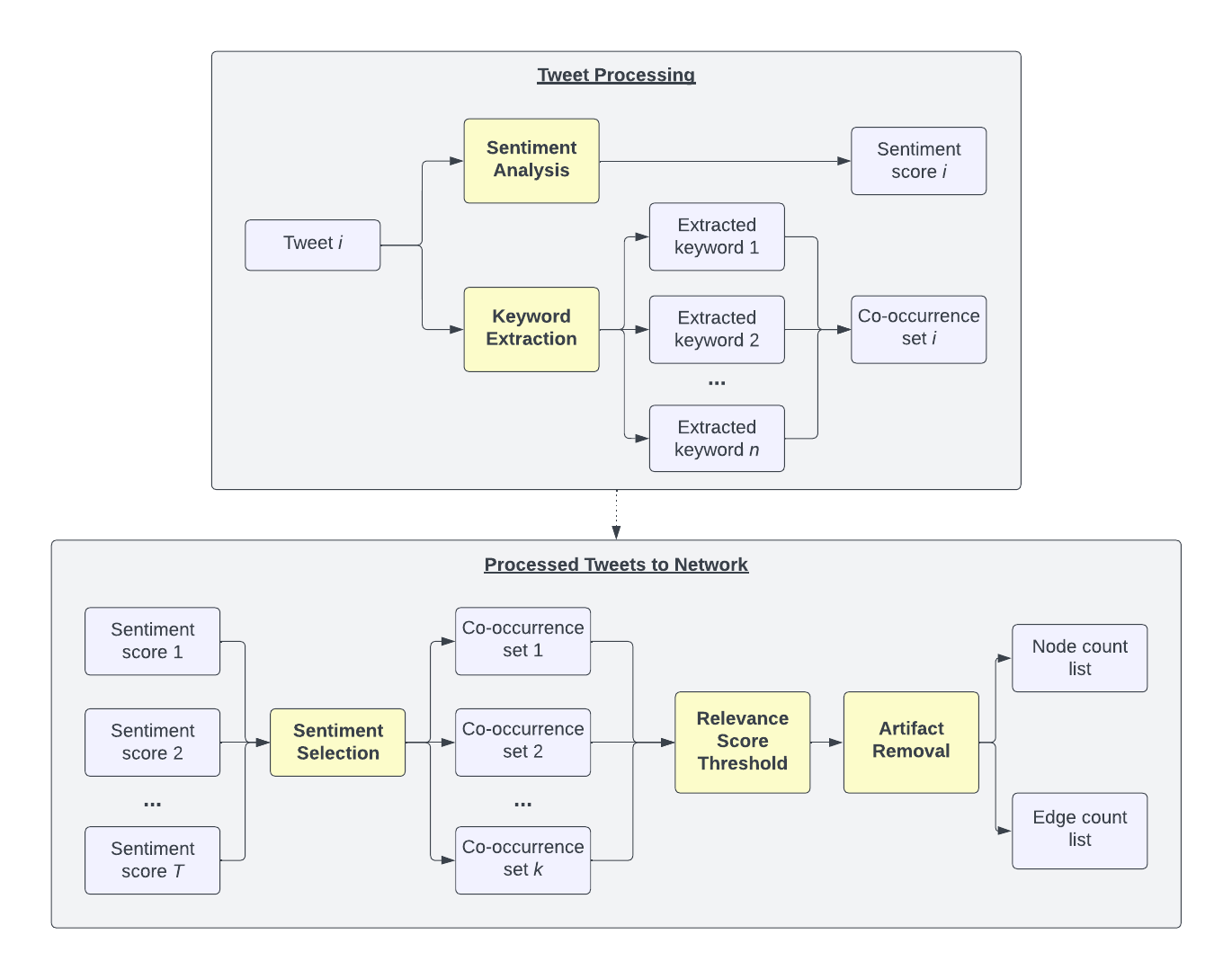}
    \caption{General methodology for processing tweets into node count and edge list. All blocks highlighted in yellow represent preprocessing decisions. The methodology is similar for Scopus articles, excluding the sentiment steps and relevance score threshold.}
    \label{fig:twitter_methodology}
\end{figure}

Figure \ref{fig:twitter_methodology} shows the general approach for the social media (Twitter) data. The bibliometric (Scopus) data are processed in a similar way.

\newpage
\subsection{Unstructured data processing}
\label{sec:data_processing}

The processing of the unstructured text consists of:
\begin{enumerate}
    \item Simultaneous extraction of keywords and their relevance scores using Rapid Automatic Keyword Extraction (RAKE) \citep{rose2010automatic};
    \item For tweet data, assignment of sentiment to the tweet using the Valence Aware Dictionary and sEntiment Reasoner (VADER) \citep{hutto2014vader}.
\end{enumerate}

For the keyword extraction, we use the NLTK \citep{bird2009natural} implementations of RAKE. RAKE is ``an unsupervised, domain-independent, and language-independent method for extracting keywords." The number of words that make up a keyword is referred to as the keyword`s length, e.g. the keyword ``Alice" has length one while ``Alice and Bob" has length 3. In order to apply RAKE, users define the possible number of words that can make up a keyword. This represents the first data preprocessing decision we analyse: we apply six different keyword length settings: 1-1, 1-2, 1-3, 2-2, 2-3, and 3-3, where a setting of $a-b$ means the shortest keywords extracted should have length $a$ (i.e. be made up of $a$ words), while the longest keywords extracted should have length $b$ (i.e. made up of $b$ words).

The relevance score from RAKE indicates how important each keyword is to the content of piece of text. Relevance scores for the above length settings can range from 1.0 to 9.0, and usually (but not strictly) correlate to the length of the keyword. 

For this analysis, we only use this simple tool for two reasons: 
\begin{enumerate}
    \item We need to repeat the keyword extraction multiple times for different parameter settings. In order to maintain reasonable running times, a fast tool is necessary.
    \item Many researchers working in these domains (particularly in the bibliometric domain) are not experienced with using complex tools, and hence a simpler tool works well as a proof-of-concept.
\end{enumerate} 
As our emphasis is on the changes that can occur in results (rather than in the results themselves), RAKE serves our purposes well, despite its shortcomings due to simplicity.

The VADER tool used for sentiment analysis on tweets is specifically intended for social media data, and gives each tweet a positive, negative, neutral, and compound sentiment score between 0 and 1. We use the standard threshold of $>0.05$ compound score for positive sentiment and $<-0.05$ compound score for negative sentiment \citep{hutto2014vader}. Tweets with scores between these values are considered neutral. This choice of tool and threshold also represents a data preprocessing decision, but we do not analyse the effect of choosing a different tool in this paper. Sentiment analysis is not relevant for the neutral Scopus data.

\begin{figure}[!h]
    \centering
    \includegraphics[width=\textwidth]{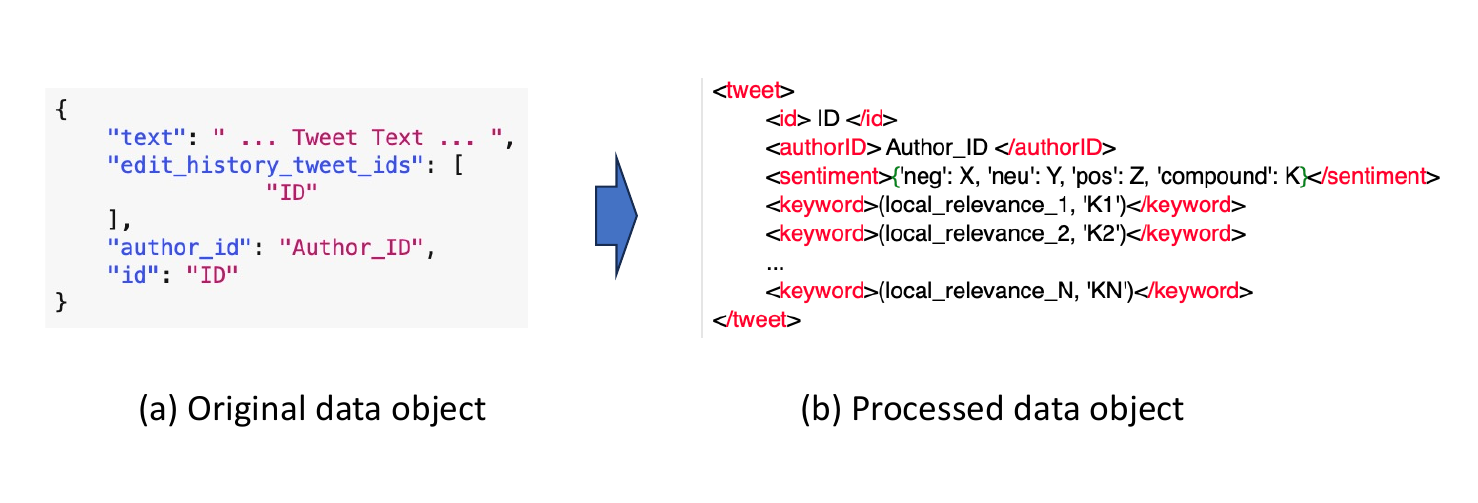}
    \caption{Original (a) and processed (b) data object.}
    \label{fig:originalVSprocessed}
\end{figure}

Figure~\ref{fig:originalVSprocessed} shows a comparison of the original data object with the one processed for tweets, with a similar process applied to articles. As shown, the original text is not included in the processed object, as it is replaced by a number of keywords. These processed objects are used to create the networks (see Section \ref{sec:engineering} and Figure \ref{fig:co_occ_abs}).

\subsection{Complex network creation}\label{sec:engineering}

\begin{figure}
    \centering
    \includegraphics[width=0.5\textwidth]{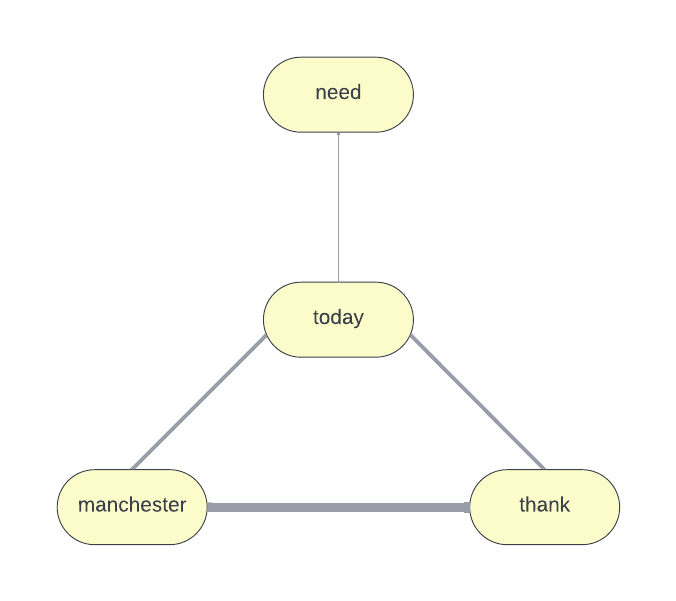}
    \caption{Example subgraph of word co-occurrence network from Twitter data, where extracted keywords form nodes and the strength of the connections between nodes (thickness of edges) are based on the frequency of co-occurrence in the same tweet}
    \label{fig:co_occ_abs}
\end{figure}

In the word co-occurrence networks, each node is an extracted keyword. The nodes are connected by weighted edges, which show how often they occur together in the same tweet or article. We also assign each node a count measure, based on how frequently the keyword occurs. Figure \ref{fig:co_occ_abs} shows how such a network might look, using a subgraph of one of the networks created using Twitter data.

For both datasets, we perform two additional processing steps during network construction:
\begin{enumerate}
    \item Shorter keywords that are contained within longer keywords are removed. For example, suppose the keywords `make America great' and `America' are both simultaneously extracted from the same tweet. We exclude the shorter keyword, `America', from all analysis for this particular tweet, i.e. this tweet will not add to the `America' node count on the network, nor add edge counts between `America' and the other keywords within this tweet.
    \item We remove artifacts of the two platforms. These are prominent keywords that occur due to the platforms themselves but have little semantic meaning. For the Twitter data, these include keywords such as `rt' (retweet), `amp' (\&), `https' etc. For the Scopus data, these are keywords related to copyright information. We remove the term `rights reserved' and any keyword containing `©' or `copyright'. Removing these keywords removes outliers from various distributions, but otherwise does not significantly affect the shape of the distribution.
\end{enumerate}

\noindent
Further, different extra steps are taken in the Twitter and Scopus data.

\subsubsection{Twitter data}

In the social media network construction, we differentiate between networks created using tweets of different sentiment. The set of tweets is divided into positive, negative, neutral, and all tweets subsets based on the VADER score.

For single occurrence keywords, we consider adding a threshold to include them within the network. For all nodes with a single count, we track the local relevance score of the keyword within its tweet. For a number of different thresholds, we create multiple networks in which keywords scoring at or below a certain relevance are removed from the network. This step is not relevant for the Scopus data, as scalability concerns already force us to remove keywords that don't occur frequently.

Multiple networks are created for the 1-1, 1-2, 1-3, 2-2, 2-3, and 3–3 keywords. For each, the networks are created using positive, negative, neutral, and all tweets. Furthermore, different relevance thresholds for removing single occurrence keywords are tested, using thresholds of 0 (no threshold), 1.0, and 4.0 being applied. These values are chosen as almost all tweets have relevance scores of 1.0, 4.0, or 9.0.

\subsubsection{Scopus data}

The number of keywords extracted from the titles and abstracts is too large for direct computation. We thus use a subset of the keywords, which needs to be selected prior to calculating co-occurrences. We only include keywords that occur in a minimum of 5 articles i.e. 0.02\% of articles. This results in networks of between 10000 and 25000 keywords, similar in size as seen in  Fu and Waltman \cite{fu2022large}. The one exception is the network created using 3-3 keywords, which is significantly smaller.

\subsubsection{Possible extensions}

It is possible to create other types of complex networks based on extracted keywords. In this paper, we focus on word co-occurrence networks, but other networks based on authors or posts as nodes can be constructed. Appendix \ref{sec:twitter_author_networks} shows an example of creating author networks based on the Twitter data, and the consequences of different data preprocessing decisions for this other type of network.

\subsection{Complex network analysis}\label{sec:analysisMethod}

We analyse the created networks at both a global scale in terms of network properties and their distributions, and at a node level in terms of node ranking.

For network properties, we replicate a similar analysis done by Fudolig \emph{et al.} \cite{fudolig2022sentiment} and Fu and Waltman \cite{fu2022large}. We compare networks on:
\begin{itemize}
    \item Number of nodes
    \item Number of edges
    \item Size of largest component
    \item Distribution of node strength
    \item Distribution of degree centrality
\end{itemize}

These node centrality measures can also be used to identify important nodes within the network. Although there are known limitations when applying these measures for ranking nodes' importance \citep{chen2012identifying, lu2016vital}, they are simple in both computational cost and their interpretation, and thus function well for gaining a basic overview. Normally, key nodes will be manually analysed in the network. In order to compare the top nodes quantitatively, the main comparison we make is in the similarity in the sorted lists of ranked nodes. These lists are compared using \emph{rank-based overlap} \citep{webber2010similarity}, a metric that measures the similarity of the ranked lists, applying a weight to indicate the relative importance of higher rated items. Interpreting this metric can be difficult, but we can generally distinguish between high similarity ($>0.8$), medium similarity ($0.8 > x > 0.3$), and low similarity ($<0.3$). High similarity in this case means that the sets of important nodes in two networks have a large overlap.

As the Scopus data are more topic-focused, it is also reasonable to perform community analysis. Such analysis is frequently performed in this domain \citep{fu2022large}. Communities are detected using the Louvain algorithm, and the top nodes are compared. In the Twitter data, it is difficult to discern any meaningful community structure, so we omit such analysis.

Due to the large number of networks considered and statistics analysed, we only report a subset of the results, especially when they offer similar insights with different setups. For example, we only compare statistics when using all tweets if the positive, negative, and neutral subsets show the same behaviour. As the focus is on highlighting the sensitivity that can be observed to these data preprocessing decisions, we also only briefly mention some results in the text. All analysis is available in two Jupyter notebooks at: \url{https://github.com/jim-g-n/twitter_scopus_data_prepro}. There are also further metrics that can be calculated for this data, but we focus on those most commonly used in the literature.

\section{Results}\label{sec:results}

This section presents the results of the experiments for the Twitter (Section \ref{sec:result_co_occ_twitter}) and Scopus data (Section \ref{sec:result_co_occ_scopus}). For both datasets, global network properties and node rankings are presented. The Scopus section additionally includes some community analysis. As there are many more data preprocessing decisions in the Twitter dataset, its results are relatively longer than the Scopus dataset.

\subsection{Twitter data}\label{sec:result_co_occ_twitter}

For the Twitter data, we present results showing first the effects of choice of keyword length and second application of a relevance score threshold on the created networks.

\subsubsection{Global network properties}

Table \ref{tab:twitter_no_thresh_prop} shows the effects of the choice of keyword length on the properties of the different networks created when no relevance score threshold is applied. Comparing the size of the largest component to the number of nodes in the network, we note that length one keywords are necessary for maintaining interpretable network structure. With this in mind, we limit our shown results to networks created using 1-1, 1-2, or 1-3 keywords for the remainder of the Twitter data analysis.

\begin{table}
\centering
\caption{Twitter Data Word Co-occurrence Network Properties (No Threshold)}
\begin{tabular}{||c c c c||} 
 \hline
 Setup & Num Nodes & Num Edges & Largest Component \\ [0.5ex] 
 \hline\hline
 Positive 1-1 & 4949 & 21832 & 4433 \\ 
 \hline
 Positive 1-2 & 8673 & 43774 & 7886 \\
 \hline
 Positive 1-3 & 10262 & 53245 & 9291\\
 \hline
 Positive 2-2 & 3724 & 3870 & 452 \\ 
 \hline
 Positive 2-3 & 5313 & 7201 & 741 \\
 \hline
 Positive 3-3 & 1589 & 485 & 4\\
 \hline
 Negative 1-1 & 4003 & 16150 & 3342 \\ 
 \hline
 Negative 1-2 & 6981 & 33248 & 6069 \\
 \hline
 Negative 1-3 & 8322 & 41054 & 7199\\
 \hline
 Negative 2-2 & 2978 & 3253 & 406 \\ 
 \hline
 Negative 2-3 & 4319 & 6078 & 719 \\
 \hline
 Negative 3-3 & 1341 & 453 & 5\\
 \hline
 Neutral 1-1 & 6848 & 17586 & 4543 \\ 
 \hline
 Neutral 1-2 & 10943 & 35589 & 7899 \\
 \hline
 Neutral 1-3 & 13301 & 45615 & 9373\\
 \hline
 Neutral 2-2 & 4095 & 3248 & 267 \\ 
 \hline
 Neutral 2-3 & 6453 & 7004 & 483 \\
 \hline
 Neutral 3-3 & 2358 & 733 & 5\\
 \hline
 All 1-1 & 12331 & 54100 & 10141 \\ 
 \hline
 All 1-2 & 22629 & 110919 & 19513 \\
 \hline
 All 1-3 & 27859 & 138216 & 23803\\
 \hline
 All 2-2 & 10298 & 10353 & 2077 \\ 
 \hline
 All 2-3 & 15528 & 20260 & 3283 \\
 \hline
 All 3-3 & 5230 & 1670 & 7\\ [1ex] 
 \hline
\end{tabular}
\label{tab:twitter_no_thresh_prop}
\end{table}

\begin{figure}
\centering
\includegraphics[width=0.9\textwidth]{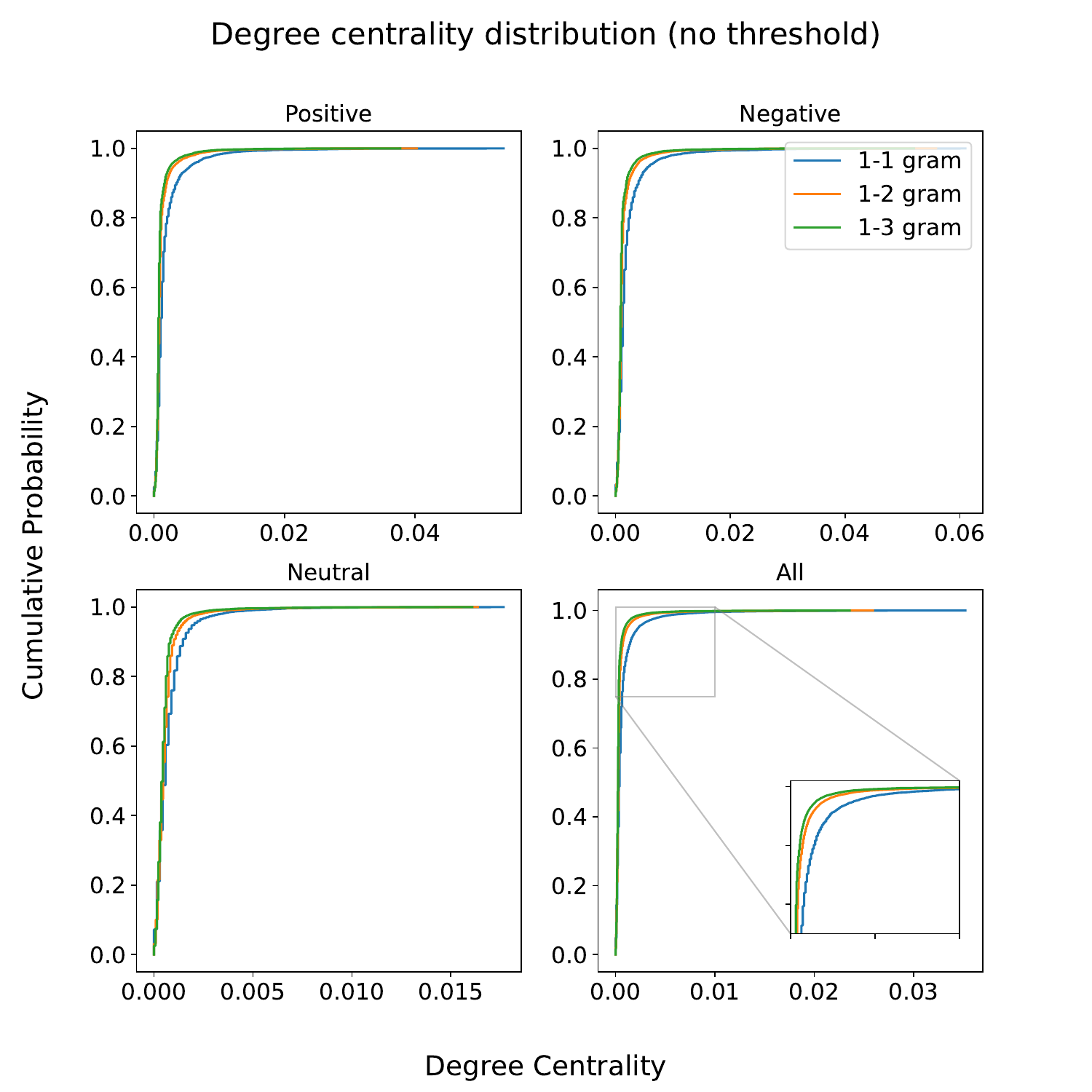}
\caption{Cumulative degree centrality distribution for word co-occurrence networks created using different length keywords, divided by positive, negative, neutral, and all tweets with no relevance score threshold.}
\label{fig:deg_cent_no_thresh}
\end{figure}

Continuing the results on the effects of choice of keyword length, Figure \ref{fig:deg_cent_no_thresh} shows the cumulative distribution function of the normalised degree centrality in the different 1-1, 1-2, and 1-3 networks when no score threshold is applied. We once again make a quick note that the results are similar between different sentiments, and so we present the results for the effect of the relevance score threshold only on the networks created with all tweets, since the impact is the same for other sentiments.

Implementing a relevance score threshold on single occurrence keywords is a way of excluding many periphery length one keywords while maintaining network connectivity. Table \ref{tab:diff_score_thresh} shows the network properties of the 1-1, 1-2, and 1-3 graphs created using all tweets and a low relevance score threshold (1.0) or a high score threshold (4.0).

\begin{table}
\centering
\caption{Word Co-occurrence Network Properties (Different Relevance Thresholds)}
\begin{tabular}{||c c c c||} 
 \hline
 Setup & Num Nodes & Num Edges & Largest Component \\ [0.5ex] 
 \hline\hline
 All 1-1 (low threshold) & 4362 & 30094 & 4048 \\ 
 \hline
 All 1-1 (high threshold) & 4362 & 30094 & 4048 \\
 \hline
 All 1-2 (low threshold) & 14690 & 74367 & 13154 \\
 \hline
 All 1-2 (high threshold) & 5662 & 41315 & 5314 \\
 \hline
 All 1-3 (low threshold) & 19957 & 96573 & 17481 \\
 \hline
 All 1-3 (high threshold) & 10929 & 57650 & 9650 \\ [1ex] 
 \hline
\end{tabular}
\label{tab:diff_score_thresh}
\end{table}

\begin{figure}
\centering
\includegraphics[width=0.9\textwidth]{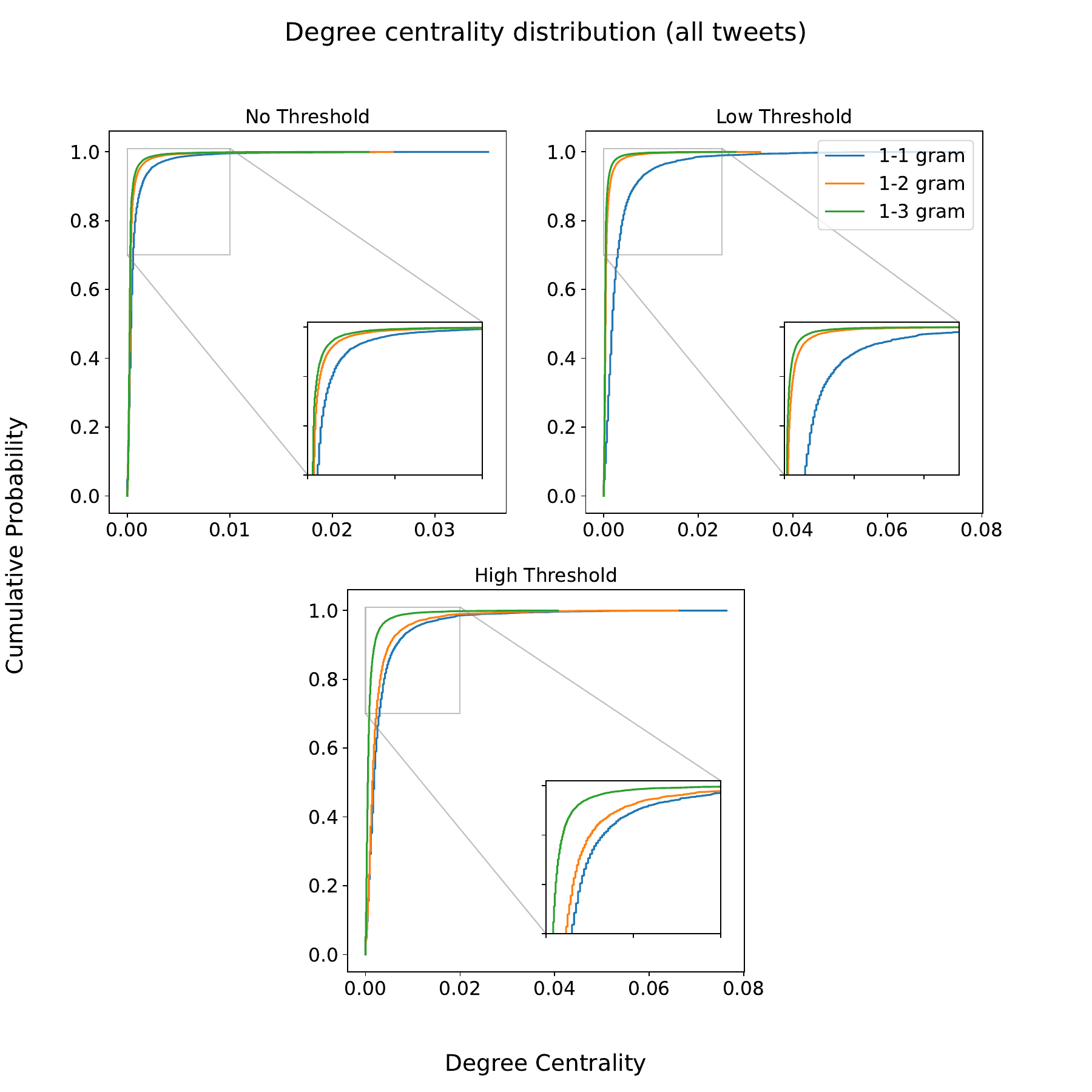}
\caption{Cumulative degree centrality distribution for word co-occurrence networks created using different length keywords using all tweets with no, low, and high relevance score threshold.}
\label{fig:deg_cent_low_thresh}
\end{figure}

To further test the interaction of keyword length and relevance score threshold on the networks, Figure \ref{fig:deg_cent_low_thresh} shows the cumulative distribution function of the normalised degree centrality with no threshold and after introducing a low (1.0) and high (4.0) threshold.

\begin{figure}
\centering
\includegraphics[width=0.9\textwidth]{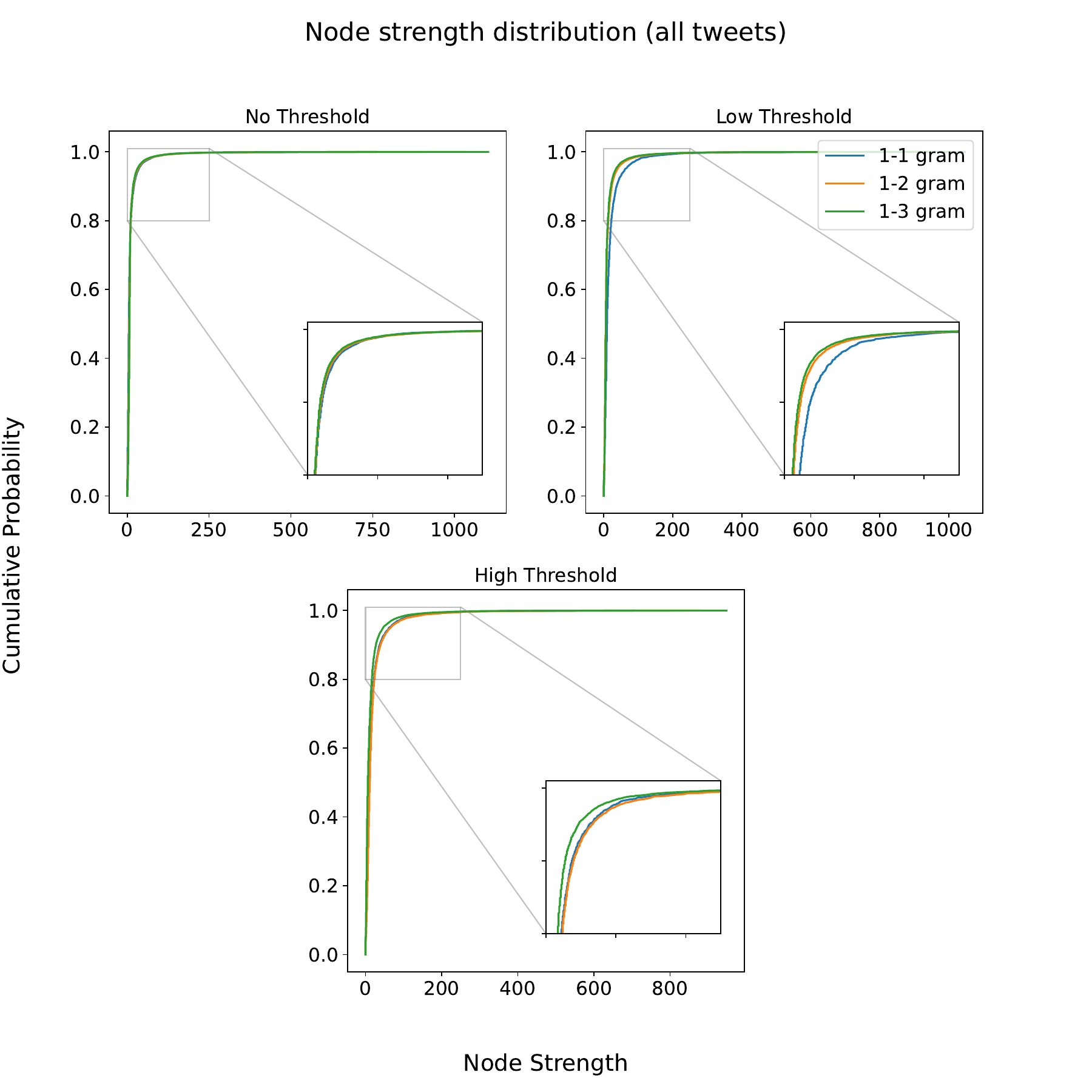}
\caption{Cumulative node strength distribution for word co-occurrence networks created using different length keywords and all tweets with no, low, or high relevance score threshold applied.}
\label{fig:node_str_diff_thresh}
\end{figure}

As a final comparison, Figure \ref{fig:node_str_diff_thresh} shows the cumulative distribution function of node strength in the 1-1, 1-2, and 1-3 networks created with all tweets when applying no, low, and high relevance score threshold. This metric scales the degree centrality by accounting for the weight of the edges between the nodes. 

\subsubsection{Node ranking}

In order to test the effects of keyword length and relevance score threshold at a more granular level, we consider node rankings. Using the normalised degree centrality or node strength of the nodes in the various networks can give us an indication of the importance of the different keywords within the networks. Unlike in the distributions, results for rankings based on degree centrality and node strength are essentially the same. With this in mind, we only report the node strength ranking similarity results. While there are some slight differences in sensitivities for networks created using different sentiments, we focus particularly on the networks created using positive or negative tweets, as these rankings could be used in practice to compare the similarity of topics discussed with different sentiment.

We compare rankings of keyword importance between a number of different networks by calculating their Rank Biased Overlap (RBO) \citep{webber2010similarity} scores using p-values of 0.9 and 0.99. A p-value of 0.9 means the highest 13 items in the list account for approximately 90\% of the similarity score, while a p-value of 0.99 means the highest 130 ranked items in the list account for 90\% of the similarity score. In other words, the p-value of 0.9 compares ranked similarity on primarily the highest ranked items, while the p-value of 0.99 compares on a much larger subset of the list.

Tables \ref{tab:node_rank_pos_0.9} and \ref{tab:node_rank_pos_0.99} show, respectively, the RBO similarity in ranked list similarity for positive networks using p-values of 0.9 and 0.99; Tables \ref{tab:node_rank_neg_0.9} and \ref{tab:node_rank_neg_0.99} show the RBO scores for negative tweets using a p-value of 0.9 and 0.99, respectively.

    \begin{table}
    \caption{Node RBO similarity based on node strength for word co-occurrence networks created using positive tweets (p-value = 0.9)}
    \setlength\tabcolsep{0pt}
\begin{tabular*}{\linewidth}{@{\extracolsep{\fill}}
                                l
                           *{9}{c}
                            }
    \toprule
                & (1)   & (2)   & (3) & (4) & (5) & (6) & (7) & (8) & (9) \\
    \midrule
(1): 1-1 no threshold    &1.000&0.956&0.948&0.881&0.891&0.912&0.881&0.855&0.878\\
(2): 1-2 no threshold     & &1.000&0.980&0.847&0.883&0.917&0.847&0.848&0.885\\
(3): 1-3 no threshold   & & &1.000&0.843&0.878&0.922&0.843&0.847&0.892\\
    \addlinespace
(4): 1-1 low threshold   & & &&1.000&0.944&0.861&1.000&0.919&0.845\\
(5): 1-2 low threshold       & & &&&1.000&0.898&0.944&0.940&0.865\\
(6): 1-3 low threshold   & & &&&&1.000&0.861&0.861&0.939\\
    \addlinespace
(7): 1-1 high threshold    & & &&&&&1.000&0.919&0.845\\
(8): 1-2 high threshold       & & &&&&&&1.000&0.900\\
(9): 1-3 high threshold   & & &&&&&&&1.000\\
    \bottomrule
\end{tabular*}
\label{tab:node_rank_pos_0.9}
    \end{table}

    \begin{table}
    \caption{Node RBO similarity based on node strength for word co-occurrence networks created using positive tweets (p-value = 0.99)}
    \setlength\tabcolsep{0pt}
\begin{tabular*}{\linewidth}{@{\extracolsep{\fill}}
                                l
                           *{9}{c}
                            }
    \toprule
                & (1)   & (2)   & (3) & (4) & (5) & (6) & (7) & (8) & (9) \\
    \midrule
(1): 1-1 no threshold    &1.000&0.902&0.889&0.900&0.876&0.867&0.900&0.825&0.825\\
(2): 1-2 no threshold     & &1.000&0.964&0.847&0.926&0.924&0.847&0.851&0.858\\
(3): 1-3 no threshold   & & &1.000&0.841&0.916&0.935&0.841&0.848&0.870\\
    \addlinespace
(4): 1-1 low threshold   & & &&1.000&0.885&0.862&1.000&0.883&0.861\\
(5): 1-2 low threshold       & & &&&1.000&0.948&0.885&0.906&0.893\\
(6): 1-3 low threshold   & & &&&&1.000&0.862&0.886&0.914\\
    \addlinespace
(7): 1-1 high threshold    & & &&&&&1.000&0.883&0.861\\
(8): 1-2 high threshold       & & &&&&&&1.000&0.936\\
(9): 1-3 high threshold   & & &&&&&&&1.000\\
    \bottomrule
\end{tabular*}
\label{tab:node_rank_pos_0.99}
    \end{table}

    \begin{table}
    \caption{Node RBO similarity based on node strength for word co-occurrence networks created using negative tweets (p-value = 0.9)}
    \setlength\tabcolsep{0pt}
\begin{tabular*}{\linewidth}{@{\extracolsep{\fill}}
                                l
                           *{9}{c}
                            }
    \toprule
                & (1)   & (2)   & (3) & (4) & (5) & (6) & (7) & (8) & (9) \\
    \midrule
(1): 1-1 no threshold    &1.000&0.942&0.945&0.951&0.912&0.924&0.951&0.868&0.889\\
(2): 1-2 no threshold     & &1.000&0.973&0.952&0.962&0.968&0.952&0.913&0.938\\
(3): 1-3 no threshold   & & &1.000&0.944&0.947&0.967&0.944&0.898&0.924\\
    \addlinespace
(4): 1-1 low threshold   & & &&1.000&0.938&0.946&1.000&0.900&0.923\\
(5): 1-2 low threshold       & & &&&1.000&0.974&0.938&0.949&0.975\\
(6): 1-3 low threshold   & & &&&&1.000&0.946&0.929&0.956\\
    \addlinespace
(7): 1-1 high threshold    & & &&&&&1.000&0.900&0.923\\
(8): 1-2 high threshold       & & &&&&&&1.000&0.970\\
(9): 1-3 high threshold   & & &&&&&&&1.000\\
    \bottomrule
\end{tabular*}
\label{tab:node_rank_neg_0.9}
    \end{table}

    \begin{table}
    \caption{Node RBO similarity based on node strength for word co-occurrence networks created using negative tweets (p-value = 0.99)}
    \setlength\tabcolsep{0pt}
\begin{tabular*}{\linewidth}{@{\extracolsep{\fill}}
                                l
                           *{9}{c}
                            }
    \toprule
                & (1)   & (2)   & (3) & (4) & (5) & (6) & (7) & (8) & (9) \\
    \midrule
(1): 1-1 no threshold    &1.000 &0.849 &0.839&0.918&0.818&0.815&0.918&0.781&0.774\\
(2): 1-2 no threshold     & &1.000 &0.959&0.834&0.934&0.939&0.834&0.865&0.878\\
(3): 1-3 no threshold   & & &1.000&0.817&0.914&0.942&0.817&0.847&0.875\\
    \addlinespace
(4): 1-1 low threshold   & & &&1.000&0.837&0.825&1.000&0.811&0.803\\
(5): 1-2 low threshold       & & &&&1.000&0.951&0.837&0.919&0.929\\
(6): 1-3 low threshold   & & &&&&1.000&0.825&0.891&0.922\\
    \addlinespace
(7): 1-1 high threshold    & & &&&&&1.000&0.811&0.803\\
(8): 1-2 high threshold       & & &&&&&&1.000&0.942\\
(9): 1-3 high threshold   & & &&&&&&&1.000\\
    \bottomrule
\end{tabular*}
\label{tab:node_rank_neg_0.99}
    \end{table}

\newpage

\subsection{Scopus data}\label{sec:result_co_occ_scopus}

For the Scopus data, we present results showing the effects of choice of keyword length.

\subsubsection{Global network properties}

Table \ref{tab:scopus_no_thresh_prop} shows the properties of the networks created using various keyword lengths. Unlike in the Twitter data, we no longer need length one keywords to maintain network structure, and so present further results for all keyword lengths.

\begin{table}
\centering
\caption{Scopus Data Word Co-occurrence Network Properties}
\begin{tabular}{||c c c c||} 
 \hline
 Setup & Num Nodes & Num Edges & Largest Component \\ [0.5ex] 
 \hline\hline
 1-1 & 10165 & 4736363 & 10156 \\ 
 \hline
 1-2 & 23645 & 10734319 & 23620 \\
 \hline
 1-3 & 25521 & 11636306 & 25490\\
 \hline
 2-2 & 13480 & 1179327 & 13480 \\ 
 \hline
 2-3 & 15356 & 1500226 & 15356 \\
 \hline
 3-3 & 1876 & 23897 & 1874\\ [1ex] 
 \hline
\end{tabular}
\label{tab:scopus_no_thresh_prop}
\end{table}

\begin{figure}
\centering
\includegraphics[width=0.8\textwidth]{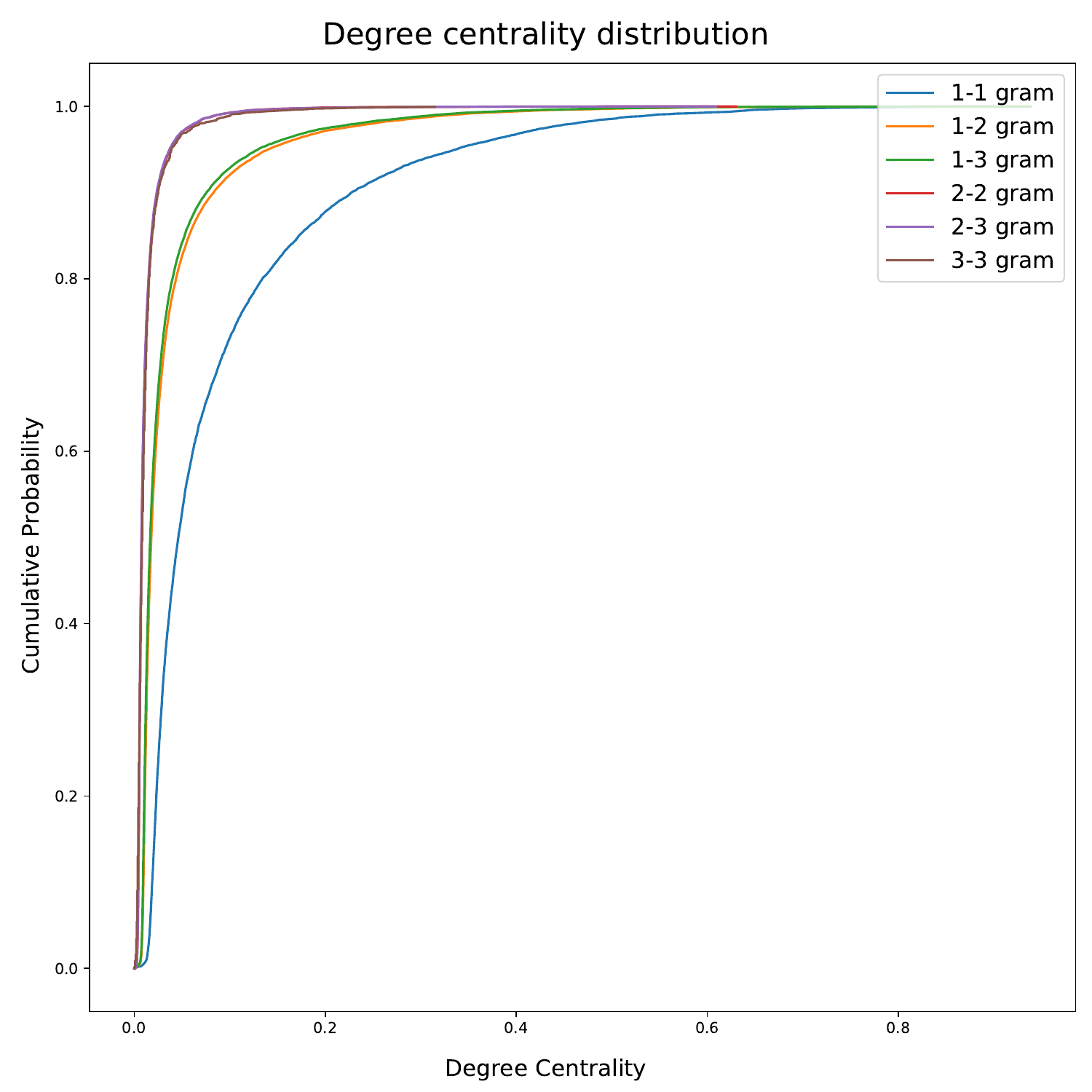}
\caption{Cumulative degree centrality distribution for Scopus word co-occurrence networks created using different length keywords.}
\label{fig:deg_cent_scopus}
\end{figure}

Due to the large differences in sizes of the networks, it is difficult to directly compare the non-normalised node strength distribution, and so we focus on the degree centralities for the Scopus data. Figure \ref{fig:deg_cent_scopus} shows the cumulative distribution of the normalised degree centralities for the different keyword length setups

\subsubsection{Node ranking}

We use the same approach as in the Twitter data to investigate the effect of keyword length on node rankings in the various networks. As before, the results are similar for degree centrality and node strength rankings, and so we present here only the node strength rankings similarity results.

Tables \ref{tab:scopus_node_rank_0.9} and \ref{tab:scopus_node_rank_0.99} show the RBO similarity in ranked list similarity for Scopus networks using a p-values of 0.9 and 0.99, respectively.

    \begin{table}
    \caption{Node RBO similarity based on node strength for word co-occurrence networks created using Scopus data (p-value = 0.9)}
    \setlength\tabcolsep{0pt}
\begin{tabular*}{\linewidth}{@{\extracolsep{\fill}}
                                l
                           *{6}{c}
                            }
    \toprule
                & (1)   & (2)   & (3) & (4) & (5) & (6)\\
    \midrule
(1): 1-1    &1.000&0.643&0.644&0.000&0.000&0.000\\
(2): 1-2     & &1.000&0.995&0.273&0.273&0.000\\
(3): 1-3   & & &1.000&0.271&0.271&0.000\\
(4): 2-2   & & &&1.000&0.891&0.000\\
(5): 2-3       & & &&&1.000&0.096\\
(6): 3-3   & & &&&&1.000\\
    \bottomrule
\end{tabular*}
\label{tab:scopus_node_rank_0.9}
    \end{table}

\subsubsection{Community structure}

Changing keyword length has effects on the community structure of the word co-occurrence networks in Scopus data. For conciseness, we focus only on the networks that include length 3 keywords -- the 1-3, 2-3, and 3-3 networks. For each, we show the largest communities (covering more than 90\% of nodes) detected using the Louvain algorithm based on edge weights. For each community, we show the size of the community and the top 5 nodes in the community based on node strength.

Tables \ref{tab:scopus_1_3_communities}, \ref{tab:scopus_2_3_communities}, and \ref{tab:scopus_3_3_communities} show the community structure of the 1-3, 2-3, and 3-3 networks, respectively.

    \begin{landscape}
        \centering 

            \begin{table}
    \caption{Node RBO similarity based on node strength for word co-occurrence networks created using Scopus data (p-value = 0.99)}
    \setlength\tabcolsep{0pt}
\begin{tabular*}{\linewidth}{@{\extracolsep{\fill}}
                                l
                           *{6}{c}
                            }
    \toprule
                & (1)   & (2)   & (3) & (4) & (5) & (6)\\
    \midrule
(1): 1-1    &1.000&0.840&0.830&0.000&0.000&0.000\\
(2): 1-2     & &1.000&0.984&0.078&0.078&0.000\\
(3): 1-3   & & &1.000&0.078&0.081&0.003\\
(4): 2-2   & & &&1.000&0.817&0.000\\
(5): 2-3       & & &&&1.000&0.181\\
(6): 3-3   & & &&&&1.000\\
    \bottomrule
\end{tabular*}
\label{tab:scopus_node_rank_0.99}
    \end{table}

    \begin{table}
    \caption{Community sizes and top 5 nodes from Louvain algorithm for network creating using 1-3 keywords}
    \setlength\tabcolsep{0pt}
\begin{tabular*}{\linewidth}{@{\extracolsep{\fill}}
                                l
                           *{4}{c}
                            }
    \toprule
                & Community 1   & Community 2   & Community 3 & Community 4\\
    \midrule
Community size    &8163&7251&5420&4656\\
Top nodes     & `climate change' &`used'&`however'&`long'\\
   & `well' & `study' &`effects'&'time'\\
   & `based' & `found' &`response'&`3'\\
       & `global warming' & `results' &`understanding'&`present'\\
   & `order' & `changes' &`likely'&`high'\\
    \bottomrule
\end{tabular*}
\label{tab:scopus_1_3_communities}
    \end{table}
    
        
        \begin{table}
        \caption{Community sizes and top 5 nodes from Louvain algorithm for network creating using 2-3 keywords}
    \setlength\tabcolsep{0pt}
\begin{tabular*}{\linewidth}{@{\extracolsep{\fill}}
                                l
                           *{5}{c}
                            }
    \toprule
                & Community 1   & Community 2   & Community 3 & Community 4 & Community 5\\
    \midrule
Community size    &4392&3609&3567&2005&1753\\
Top nodes     & `climate change' &`climate variability'&`results suggest'&`results show'&'results showed'\\
   & `global warming' & `american geophysical union' &'global climate change'&`results indicate'&'verlag berlin heidelberg'\\
   & `case study' & `climate changes' &'climate warming'&`study area'&'important role'\\
       & `springer science' & `intergovernmental panel' &'sons ltd'&`21st century'&'growing season'\\
   & `business media b' & `climatic changes' &`changing climate'&'future climate change'&'present study'\\
    \bottomrule
\end{tabular*}
\label{tab:scopus_2_3_communities}
    \end{table}


        \begin{table}
        \caption{Community sizes and top 5 nodes from Louvain algorithm for network creating using 3-3 keywords}
\begin{subtable}{1\linewidth}
    \setlength\tabcolsep{0pt}
\begin{tabular*}{\linewidth}{@{\extracolsep{\fill}}
                                l
                           *{3}{c}
                            }
    \toprule
                & Community 1   & Community 2   & Community 3\\
    \midrule
Community size    &406&378&361\\
Top nodes     & `american geophysical union' &'business media dordrecht'&`future climate change'\\
   & `sea surface temperature' &'global warming potential'& `business media b' \\
   & `anthropogenic climate change' &'greenhouse gas emissions'& `climate change impacts' \\
       & `north atlantic oscillation' &'life cycle assessment'& `climate change scenarios' \\
   & `pacific decadal oscillation' & 'climate change mitigation'&`general circulation models' \\
    \bottomrule
\end{tabular*}
\end{subtable}
\begin{subtable}{1\linewidth}
    \setlength\tabcolsep{0pt}
\begin{tabular*}{\linewidth}{@{\extracolsep{\fill}}
                                l
                           *{3}{c}
                            }
    \toprule
                & Community 4   & Community 5   & Community 6\\
    \midrule
Community size    &225&170&137\\
Top nodes     & '5 ° c'&`last glacial maximum'&`global climate change'\\
   & '2 ° c'&`little ice age'&`verlag berlin heidelberg'\\
   & '4 ° c'&`western united states'&`sea level rise'\\
       & '3 ° c'&`western north america'&`verlag gmbh germany'\\
   & '1 ° c'&`term climate change'&`ongoing climate change'\\
    \bottomrule
\end{tabular*}
\end{subtable}
\label{tab:scopus_3_3_communities}
    \end{table}

\end{landscape}
    \clearpage

\section{Discussion}\label{sec:discussion}

\subsection{Twitter data}

\subsubsection{Global network properties}

The results of the analysis of the global network properties for the word co-occurrence networks show that, regardless of which metrics/distributions are analysed, the choice of keyword length and relevance score threshold has an effect. This effect can be seen in positive, negative, neutral, and all tweets. While there is always some effect, certain network metrics are more sensitive than others.

We begin by analysing Table \ref{tab:twitter_no_thresh_prop}. The first effect of changing keyword length that can be seen is in the number of nodes within the network. Going from 1-1 to 1-2 to 1–3 keywords increases the size of the networks, as most of the shorter keywords stay the same and longer keywords are simply added on as extra nodes within the network. Going from 1-1 to 1–3 keywords roughly doubles the number of nodes in the networks, for positive, negative, neutral, and all tweets i.e. average scaling of number of keywords from the keyword extraction algorithm is roughly the same regardless of the sentiment of the tweet.

A second effect of the length of keywords is on network connectivity in terms of largest component. For networks created with the 1-1, 1-2, or 1–3 keywords, the largest component within the networks includes the majority of nodes. This is in contrast to the networks created with 2-2, 2-3, and 3-3 length keywords. All these networks created without length one keywords suffer from a lack of connectivity. Networks created with 2-2 or 2–3 keywords have between roughly 7.5 - 20\% of nodes in the largest component of the network, and this can vary based on the sentiment of the tweets for the given keyword length. Networks created with 3–3 keywords have little discernible network structure, consisting of many small, highly connected components. This is likely the result of the characteristics of the dataset. Since it is not topic-specific, longer keywords are less likely to occur together.

Effects can further be seen in Figure \ref{fig:deg_cent_no_thresh}. Once again, keyword length has an effect on the metric, with comparable impact regardless of sentiment. The networks created with the 1-1 keywords have consistently higher degree centralities on average across all sentiments. This is as expected, since 1-1 keywords are less specific and thus more likely to occur together. The differences in distributions between 1-2 and 1-3 keyword networks are less significant than between them and 1-1 keyword networks. The distributions have similar shapes for positive, negative, neutral, and all tweets. The negative and positive networks have higher centrality on average, while the neutral network has the lowest average centrality. These first results show that the networks can be affected by the choice of keyword length. 

Adding a relevance score threshold also has noticeable effects. We begin by considering Table \ref{tab:diff_score_thresh}. As with the keyword length, the number of nodes within the network is affected by this score threshold preprocessing step. Applying the minimum relevance score threshold of 1.0 removes a significant number of the length one keywords, where the 1-1 network created from all tweets reduces in size from 12331 to 4362 nodes. Likewise, the 1-2 and 1-3 networks reduce from 22629 and 27859 nodes to 14690 and 19957 nodes, respectively. Unlike the 2-2, 2-3, and 3-3 networks from before, these relevance score threshold networks retain relatively high connectivity, with close to 90\% of nodes in the largest component for both networks. Because the connectivity of the network is not as sensitive to the relevance score threshold as to forcing longer keyword lengths, application of such a threshold is a potential way to reduce the number of short, periphery words in the network while retaining some structure. 

Increasing the threshold further reduces the number of nodes in the 1-2 and 1-3 networks, while keeping a majority in the largest component (the threshold has no further effects on the 1-1 network since all length one keywords have a relevance score of 1.0). This higher relevance score threshold can thus be used to filter out longer length keywords that also fall on the periphery of the network. However, these longer keywords may be more meaningful than the length one keywords, as they are typically more specific and hence potentially relevant for analysing the network.

Effects are also seen in degree centrality distributions, as shown in Figure \ref{fig:deg_cent_low_thresh}. In the no threshold case, the distributions of the 1-2 and 1-3 networks are very similar, while that of the 1-1 network is slightly different. Applying a low threshold amplifies this difference in distribution. The 1-1 network still has, on average, a higher degree centrality, but this difference is now more pronounced, while the 1-2 and 1-3 networks are still relatively similar. As expected, all networks show higher average degree centrality when the threshold is introduced and periphery nodes removed. This is more extreme in the 1-1 network, where the highest degree nodes are doubled in centrality, while the 1-2 and 1-3 networks show far lower increases. 

Applying a high threshold causes more significant changes. In particular, the degree centralities of the 1-2 network are now closer to those of the 1-1 network, while the 1-3 network shows smaller changes. These results show that the choice of length of keyword extraction and application of a relevance score threshold interact with each other and can have varying levels of impact.

We can compare these effects to node strength distributions (Figure \ref{fig:node_str_diff_thresh}). This metric shows less sensitivity to both the choice of keyword length and a relevance score threshold, with smaller differences in the distribution when compared to degree centrality.

With no threshold, all three networks show very similar distributions. Applying a low score threshold slightly separates the 1-1 network distribution from the 1-2 and 1-3 networks, while a high score threshold brings the 1-2 network closer to the 1-1 network with the 1-3 network remaining relatively similar. However, these changes are still significantly less pronounced than those in the degree centrality distributions. This behaviour is consistent when excluding extreme high strength outlier nodes from the plot. As a result, it may be that this metric is more robust to the choice of keyword extraction and relevance score threshold. In the case where both metrics are relevant, it could be reasonable to focus on the node strength distributions, due to their better reliability. However, it is not obvious a priori how different metrics will behave under different data preprocessing, emphasising the need to investigate different preprocessing pipelines.

While not shown, the node count distributions show similar behaviour to the degree centrality and node strength distributions. With no score threshold, the 1-1 network is slightly different, and this difference is increased when applying a low score threshold. Applying a high score threshold moves the 1-2 network to be similar to the 1-1 network while the 1-3 network remains relatively stable.


In summary, all metrics tested show some sensitivity to the choice of keyword length and relevance score threshold, and these effects can interact. These sensitivities are consistent across different sentiments. Some distributions, like node strength, seem to be less affected by these preprocessing steps, while metrics like number of nodes in the network can drastically change. It can be difficult to tell in advance which metrics are sensitive and which are not. Furthermore, in this dataset, length one keywords are crucial for maintaining network connectivity. If the objectives of the study rely on network analysis, this could dictate, to an extent, the choices that can be made with regard to keyword extraction.

\subsubsection{Node ranking}

As opposed to the global network properties, node rankings in the word co-occurrence network show relatively high robustness to the choice of keyword lengths and relevance score threshold. Ranked list similarities are generally within the 80\% -- 98\% range for both positive and negative networks. 

Looking first at Table \ref{tab:node_rank_pos_0.9}, we see that the largest similarity is generally between the non-thresholded networks and is around 0.95. The lowest similarity is between 1-1 and 1-3 networks where one has no threshold and another a high threshold, and is around 0.85. There is a similarity of 1 between the 1-1 low threshold and 1-1 high threshold networks, as these networks are identical.

We compare this to the values in Table \ref{tab:node_rank_pos_0.99}. These similarity values, where more elements within the ranked list are compared and weighted towards the score, are generally lower than those calculated using a p-value of 0.9. The most extreme difference is between the network created using 1-1 keywords and no threshold and 1-3 keywords and a high threshold. That being said, the RBO score of 0.83 is still relatively high, showing that the node rankings are reasonably robust to changes in the data preprocessing. 

These results can be compared to those found in the general network properties. The sizes of the networks created using different data preprocessing can differ drastically; however, the important nodes within the networks remain stable. This means that the data preprocessing does not drastically edit or remove important nodes. This is consistent with what was observed in the 2-2, 2-3, and 2-3 networks, where network structure broke down due to the loss of length one keywords. These length one keywords thus play an important role in network connectivity, and as such will be given high rankings in the node importance.

The results for negative tweets are relatively similar to positive tweets, and so we only highlight an interesting takeaway. Looking at Tables \ref{tab:node_rank_neg_0.9} and \ref{tab:node_rank_neg_0.99}, we note that the RBO scores for negative tweets are within similar ranges to positive tweets, but show a key difference: using a p-value of 0.9, the lowest score is 0.868, slightly higher than the lowest score of 0.843 in the positive tweets. Using a 0.99 p-value, the lowest score is 0.774 in the negative tweets, which is substantially lower than the lowest score of 0.825 in the positive tweets. Hence, the higher ranked nodes in the negative networks are more stable to changes in data preprocessing than in the positive networks, but the lower ranked nodes are less stable in the negative tweets. Whether these differences in values have a significant or practical implication would depend on the exact interpretation of the results, but nevertheless it is important that anyone analysing the results is aware of the sensitivity and can potentially assess the impact of this sensitivity.

\subsection{Scopus data}

\subsubsection{Global network properties}

As in the Twitter data, all global network properties for the word co-occurrence networks are affected by the choice of keyword length. There are, however, some noticeable differences as a result of the more focused domain from which the data are collected.

We first consider Table \ref{tab:scopus_no_thresh_prop}. As in the Twitter data, the main effect that can be seen is in the number of nodes and edges in the network. Going from 1-1 to 1-2 to 1–3 keywords generally adds more nodes and edges at each step, with roughly 2.5 times the number of nodes in the 1-3 versus the 1-1 networks. The 2-2 and 2-3 networks have sizes in between the 1-1 and 1-2 networks. The 3-3 network is significantly smaller than all other networks.

Unlike in the Twitter data, network connectivity is not affected by changing the keyword lengths. In all setups, the largest component includes 99.9\% to 100\% of nodes. This is a result of articles being collected based on related search terms. Furthermore, there is more of a consistent terminology in scientific writing compared to the informal vocabulary used in social media such as Twitter.

In Figure \ref{fig:deg_cent_scopus}, we see there a noticeable trend in the distributions. The 1-1 network has the highest degree centralities; the 1-2 and 1-3 networks have similar distributions and on average lower centralities than the 1-1 network; finally, the 2-2, 2-3, and 3-3 networks have again lower average centralities while being similar to each other. Going from the 1-2 to 1-3 network only adds a relative small number of nodes and likewise for going from 2-2 to 2-3 -- this explains why these distributions are quite similar.

The results for the Scopus data are overall comparable to the Twitter data. Once again, key metrics such as number of nodes and distributions of centrality are strongly influenced by choice of keyword length. By focusing on a specific topic, connectivity is maintained regardless of the keyword length used. 

\subsubsection{Node ranking}

Unlike in the Twitter data, node rankings in the Scopus word co-occurrence network can be highly sensitive to choice of keyword length. Another distinction is that there are now networks that have zero overlap in keyword lengths.

We begin with the RBO similarity scores shown in Table \ref{tab:scopus_node_rank_0.9}. The 0 similarity scores are between networks that have no overlap in possible keyword length. The 1-1 network has moderate similarity (0.643) to the 1-2 and 1-3 networks, which in turn have very high similarity with each other (0.995). These two networks (1-2 and 1-3) have low similarity to the 2-2 and 2-3 networks (around 0.272). The 2-2 and 2-3 networks have high similarity (0.891), while the 2-3 and 3-3 networks have very low similarity (0.096). The general trend is that networks with the same minimum length keyword will have moderate to high similarity with each other (e.g. 1-2 and 1-3 or 2-2 and 2-3). This is because the shorter keywords tend to be have high strength (and thus high ranking) in the networks.

When increasing the p-value, some similarity scores increase while others decrease, as seen in Table \ref{tab:scopus_node_rank_0.99}.  The 1-1 network is now more similar to the 1-2 and 1-3 networks than before (approximately 0.835 compared to 0.643); likewise, the 2-3 is now more similar to the 3-3 (0.181 versus 0.096). On the other hand, the 1-2 and 1-3 networks are less similar to the 2-2 and 2-3 networks (from around 0.272 before to around 0.078 now). The overall trend is the same as in the previous RBO calculation, where the minimum length keyword is the most important for similarity comparisons; however, the longer keywords now seem to play a bigger role, which can vary the similarity scores.

In comparison to the Twitter data, similarity scores are much lower for the Scopus data. While exact interpretation of these values is challenging, we can see that there can arise very large differences in top ranked/important nodes, especially when varying the minimum length keyword.

\subsubsection{Community structure}

Finally, we analyse the community structure shown in Tables \ref{tab:scopus_1_3_communities}, \ref{tab:scopus_2_3_communities}, and \ref{tab:scopus_3_3_communities}. The first difference that can be seen is in the number of communities within the networks and their relative sizes. The 1-3 network has 4 communities, varying in size between 18.2\% and 32\% of the nodes in the network. On the other hand, the 2-3 network has 5 communities each covering between 11.4\% and 28.6\% of the nodes. Finally, the 3-3 network has 6 communities, between 7.3\% and 21.6\% of the nodes. The 3-3 network has thus both the largest number of and most similarly sized communities. The 1-3 network has the fewest communities and the largest minimum size community. In general, it seems that the longer the minimum keyword length, the more communities there are, and these communities are of more similar sizes.

The second distinction is in the top nodes within each community. In all three cases, it is possible to interpret the communities to some extent. For example, in the 1-3 network, community 2 contains general research terms, which is similar to community 4 in the 2-3 network. Community 4 in the 3-3 network covers temperatures, while community 1 is generally about ocean and wind patterns. That being said, not all communities can be reasonably interpreted, such as community 4 in the 1-3 network. Overall, the 3-3 keywords (and their specificity) allow for more clear interpretation of community structure. As discussed for the Twitter data, the intended analysis could thus determine which preprocessing decisions are most reasonable -- if community analysis is crucially important, using longer keywords only makes more sense.  

A final observation is in the length of the top nodes in each community. While all three networks allow for length 3 keywords, the top nodes are usually of the minimum length for a given network. We see this, for example, in the only length 2 keywords in the 1-3 results being the very broad terms `climate change' and `global warming' -- terms that directly relate to the topic being modelled. Hence, even when the dataset is topic-focused, shorter keywords generally play a larger role in the constructed networks.

\subsection{Limitations}

As mentioned previously, there are some limitations to this study. The most prominent is in the Twitter dataset. The changes to the Twitter API severely limited the total number of tweets that could be collected. Fortunately, we were still able to collect a similar sized dataset as used in other studies. This means it was possible to identify potential risks and sensitivities, but it is difficult to draw generalisable conclusions given this sample size. That being said, this case emphasises the need to transparency in data preprocessing. It is not guaranteed that once publicly-available data will remain as such. Experiments performed using limited or black-box preprocessing approaches may become irreproducible if data availability changes.

The Scopus data analysis, on the other hand, was much larger and complete, allowing for stronger, more convincing claims. This came at the cost, however, of computational challenges. Implementing an approach as detailed in this paper would require careful optimisation when working with larger datasets, hence limiting which keyword extraction tools are feasible.

Finally, we have focused on two datasets from two different domains. While results are consistent between the two, generalising all claims would require further experimentation. However, these initial indications are strong, and align well with results for other data preprocessing decisions in complex network construction in other domains.

\section{Conclusions and future work}
Due to privacy, ownership, and analytical considerations when analysing social media or bibliometric data, complex networks are frequently used. The construction of these networks inherently requires that decisions be made. These decisions are often made on the basis of interpretability and scalability. Although both are reasonable concerns, such choices can have far-reaching impacts that can change the interpretation of results.

To illustrate this, we tested some of these decisions and the impact they can have on networks created based on word co-occurrence in Twitter and Scopus data. We find that:
\begin{itemize}
    \item Many of the standard network metrics reported in the literature are sensitive to changes in both the length of keywords extracted and, for the Twitter data, the application of some form of keyword relevance selection.
    \item These sensitivities are not always consistent between different types of data, and it can be difficult to predict a priori how they might play out.
\end{itemize}

The Twitter dataset used in this paper has some limitations due to its characteristics of small size and lack of focused topic, and some of these sensitivities may be exacerbated by these characteristics. However, the sensitivities observed are also present in the large-scale, topic-specific Scopus data. This suggests that these results could also be applicable in other domains, and such potential sensitivities should thus be taken into consideration if using automatic keyword extraction.

As a result, this analysis could just as easily be applied to other social networks or text sources where keyword extraction is employed. Doing so provides a more transparent picture of the nature of networks, allowing researchers to have greater confidence in conclusions that are consistent between different data preprocessing decisions.

\backmatter
\newpage
\section*{Declarations}

\bmhead{Data availability} Tweet data were collected using Twitter`s API (\url{https://developer.twitter.com/en/docs/twitter-api}) and the tweet IDs from the Twitter Parliamentarian Database \cite{van2020twitter}. Bibliometric data were collected using Scopus Search (\url{https://www.scopus.com/search/form.uri?display=basic}) and the search terms `climatic chang*', `climate variabilit*', `climatic variabilit*', `climate warming', `climatic warming', `climate chang*', and `global warming'.

\bmhead{Code availability} The code used for analysis is available here: \url{https://github.com/jim-g-n/twitter_scopus_data_prepro}. 


\newpage
\begin{appendices}

\section{Twitter author networks}
\label{sec:twitter_author_networks}
As an extension to the word co-occurrence networks, we present results for author networks based on the Twitter data. In this section, we offer a discussion on the results as they are presented.

\subsection{Author networks creation}

\begin{figure}
    \centering
    \includegraphics[width=0.5\textwidth]{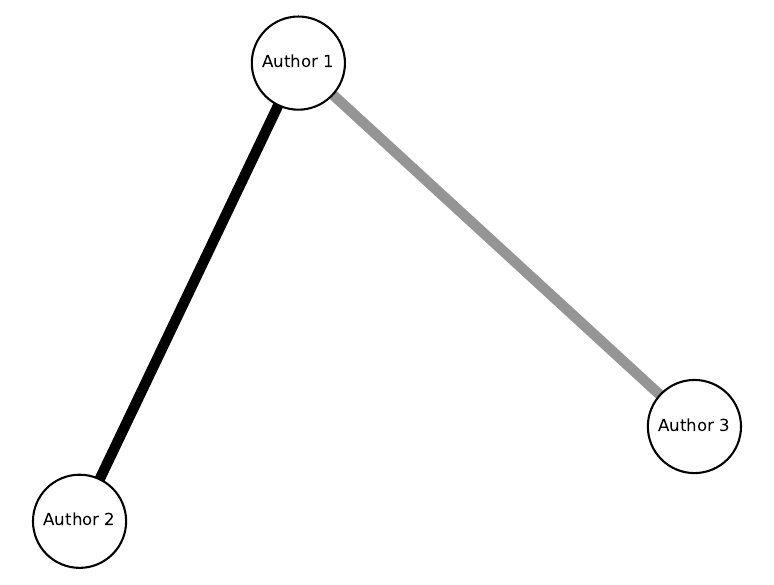}
    \caption{Author network, where authors form nodes and the strength of connections between nodes are based on similarity in keywords extracted from authors' tweets.}
    \label{fig:author_abs}
\end{figure}

In these author networks, each node is a tweet author, and the nodes are connected based on the similarity of the keywords used by each author. Figure \ref{fig:author_abs} shows an illustrative example.

For each author, we create a set of all the keywords used by them. For each pair of authors, we calculate the Jaccard similarity of the sets of keywords used by the authors. This similarity indicates the weight of the edge between the authors, with authors remaining unconnected when they have never used the same keyword.

We once again exclude keywords that are contained within larger keywords in the same tweet. However, we no longer remove single occurrence keywords. These keywords more likely form a part of the author`s focus, and are no longer periphery to the network as in the co-occurrence network. Furthermore, authors also use many more words, so the effect of these single occurrence words are less significant. 

For the author networks, we consider keeping and removing the noisy keywords. This decision also falls under data preprocessing, and has a significant impact on certain results within this network.

Similarly to the co-occurrence networks, author networks are constructed from 1-1, 1-2, 1-3, 2-2, 2-3, and 3–3 keywords collected from positive, negative, neutral, and all tweets.

\subsection{Author networks results}\label{sec:results_author}

\subsubsection{Global network properties}

We perform similar analysis of global network properties for the author networks as was done for the word co-occurrence networks. Due to the difference in nature of the author networks compared to the co-occurrence networks, the metrics that are affected and how they are affected is different. However, there are still numerous sensitivities that can be observed in relation to keyword length and keyword removal.

As in the word co-occurrence networks, the effects observed are comparable for positive, negative, neutral, and all tweets, so we limit our analysis to networks created with all tweets. Table \ref{tab:author_network_prop} shows the network properties of the author networks for both different keyword lengths and whether noisy keywords have been removed.

A first observation is the differences in the effect of keyword length on number of nodes in the network. As the set of authors does not change based on alterations to the keywords, the number of nodes in the author networks is a constant 872. This is in contrast to the high sensitivity of number of nodes to keyword length in the word co-occurrence networks. However, the number of edges and particularly their weights in the author networks can change. For example, the `all 1-1' network has a total edge weight of 10196.86, while the `all 3-3' has a total edge weight of only 84.39. In general, the weights of the edges in these networks gradually decrease as longer keywords are introduced. This is simply due to the fact that the longer keywords are more specific, and thus less likely to be used by multiple authors.

A second key contrast to the word co-occurrence networks is in the effect of keyword length on connectivity -- the author networks created with 2-2, 2-3, and 3-3 keywords are generally well-connected, with the largest component and number of total edges  similar to the 1-1, 1-2, and 1-3 networks. While there is still a slight reduction in connectivity as the minimum keyword length is increased, there are clearly some keywords, even in the 3-3 network, that are regularly used by different authors to connect them. 

We also consider the effect of removing noisy keywords. Doing so does not change the connectivity of the networks, indicating that authors are not connected exclusively through noisy keywords. However, it does reduce the weight of the edges within the networks. In the 1-1, 1-2, and 1-3 cases, total edge weight is reduced to roughly one quarter if noisy keywords are removed. In the 2-2 and 2-3 cases, it reduces to roughly half. There are no length 3 noisy keywords, so the 3-3 network is unaffected. 

\begin{table}
\centering
\caption{Author Network Properties}
\begin{tabular}{||c c c c||} 
 \hline
 Setup & Num Edges & Edge Weight & Largest Component \\ [0.5ex] 
 \hline\hline
 All 1-1 & 372816 & 10196.86 & 864 \\ 
 \hline
 All 1-2 & 377146 & 7053.18 & 869 \\
 \hline
 All 1-3 & 377146 & 6058.59 & 869\\
 \hline
 All 2-2 & 348195 & 1352.13 & 835 \\ 
 \hline
 All 2-3 & 359976 & 982.02 & 849 \\
 \hline
 All 3-3 & 267546 & 84.39 & 732\\
 \hline
 All 1-1 (noise removed) & 372816 & 2362.73 & 864 \\ 
 \hline
 All 1-2 (noise removed) & 377146 & 1766.17 & 869 \\
 \hline
 All 1-3 (noise removed) & 377146 & 1558.81 & 869\\
 \hline
 All 2-2 (noise removed) & 348195 & 596.92 & 835 \\ 
 \hline
 All 2-3 (noise removed) & 359976 & 442.91 & 849 \\
 \hline
 All 3-3 (noise removed) & 267546 & 84.39 & 732\\ [1ex] 
 \hline
\end{tabular}
\label{tab:author_network_prop}
\end{table}

Unlike in the word co-occurrence networks, the largest components within the author networks are mostly fully-connected, meaning the degree centrality distributions consist of a number of nodes with degree centrality 0 and the rest all having exactly equal centrality close to 1. This means there is a set of authors who have no keywords in common with anyone, and the remainder all have at least one keyword in common. As a result, our network analysis is limited to the node strengths, which do show differences in distributions and values.

\begin{figure}[H]
\centering
\includegraphics[width=0.9\textwidth]{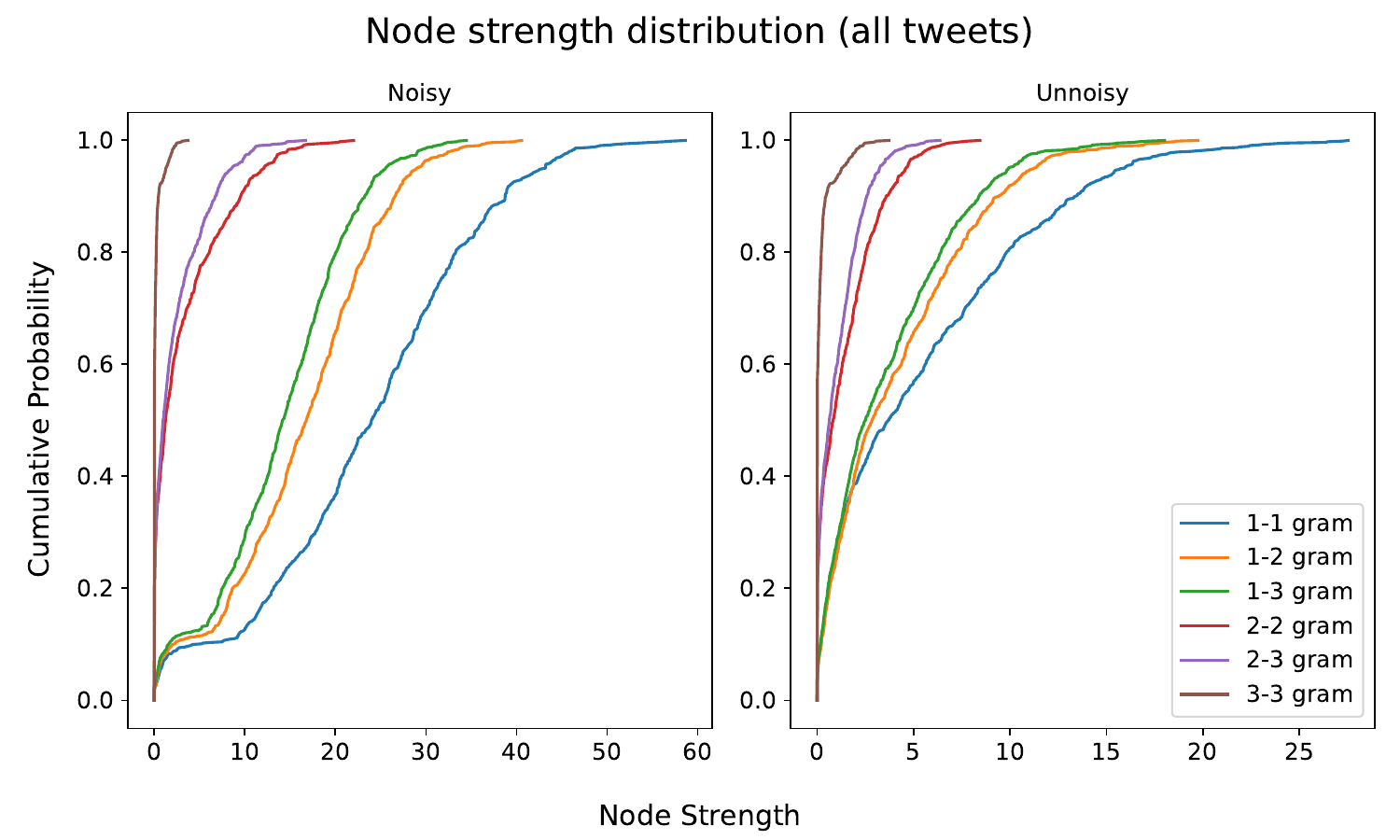}
\caption{Cumulative node strength distribution for author networks created using different length keywords and all tweets with noisy keywords kept or removed.}
\label{fig:node_str_diff_thresh_authors}
\end{figure}

Figure \ref{fig:node_str_diff_thresh_authors} shows the distributions of the node strengths for the different networks when noisy keywords are retained (noisy) or removed (unnoisy). Unlike in the word co-occurrence networks, the node strength distributions in the author networks show much higher sensitivities to the data preprocessing. As could be expected, the `less complex' the keywords in terms of length, the higher the node strengths on average. There are less pronounced differences between the 1-2 and 1-3 networks and the 2-2 and 2-3 networks than between other pairs of networks. The 1-2 and 1-3 having smaller differences is consistent with what was observed in the word co-occurrence networks, and going from 2-2 to 2-3 networks follows a similar pattern of mostly just adding length 3 keywords which play a less crucial role than the length 2 keywords.

Removing noisy keywords generally reduces the node strength. It also makes the distributions smoother, with the 1-1, 1-2, and 1-3 networks now having a similar number of low strength nodes. 

These results highlight again how the choice of data preprocessing can affect global network properties, even when the networks constructed are of a different nature (compared to the co-occurrence networks). At the same time, different global properties can be affected in different ways, such as more sensitivity in the node strength distribution. This emphasises the need to consider properties on a case-by-case basis, and that there is no `one-size-fits-all' approach.

\subsubsection{Node ranking}

As in the word co-occurrence networks, we once again rank nodes based on their node strength and apply RBO to compare the similarities of the lists in positive and negative networks.

In contrast to the word co-occurrence networks, the node rankings in the author networks show very high sensitivity to the data preprocessing, with much more extreme differences in both the highly ranked nodes and a larger subset of the nodes. The behaviour of the ranking similarity when considering more or less nodes is reversed from the co-occurrence networks, where now comparing more nodes (higher p-value) increases the ranking similarity.

\bigskip
\noindent
\textbf{Positive tweets}

Tables \ref{tab:author_rank_pos_0.9} and \ref{tab:author_rank_pos_0.99} show the RBO similarities for the positive author networks node rankings for p-values of 0.9 and 0.99, respectively. `noisy' refers to networks that still include noisy keywords, while `unnoisy' refers to those that have had noisy keywords removed. In general, the similarities observed are much lower than in the word co-occurrence networks. 

With a p-value of 0.9, the 1-1, 1-2, and 1-3 networks with noise retained show similarities between 50\% and 85\%, compared to the approximately 95\% no threshold networks in the word co-occurrence networks. Going from 1-1, 1-2, or 1-3 networks to 2-2, 2-3, or 3-3 networks drastically lowers the similarity to under 20\%, much lower than the minimum 85\% in the high threshold co-occurrence networks (which also remove short length keywords). While these approaches are not identical, they show again how slight changes in data preprocessing can have a big impact and that this is dependent on the types of network being constructed.

Using a p-value of 0.99, the similarities increase for all pairs of networks. This means that, in the author networks, the most extreme highly ranked nodes can differ substantially, but these differences do decrease as we move further down the list. This is also different to what was observed in the word co-occurrence networks.

    \begin{table}
    \caption{Node RBO similarity based on node strength for author networks created using positive tweets (p-value = 0.9)}
    \setlength\tabcolsep{0pt}
\begin{tabular*}{\linewidth}{@{\extracolsep{\fill}}
                                l
                           *{11}{c}
                            }
    \toprule
                & (1)   & (2)   & (3) & (4) & (5) & (6) & (7) & (8) & (9) & (10) & (11) \\
    \midrule
(1): 1-1 (noisy)    &1.000 &0.500 &0.501&0.004&0.004&0.009&0.005&0.023&0.031&0.043&0.058\\
(2): 1-2 (noisy)     & &1.000 &0.827&0.114&0.114&0.005&0.007&0.032&0.033&0.088&0.067\\
(3): 1-3 (noisy)   & & &1.000&0.162&0.162&0.008&0.008&0.037&0.051&0.148&0.120\\
(4): 2-2  (noisy)  & & &&1.000&0.741&0.003&0.008&0.008&0.008&0.035&0.005\\
(5): 2-3  (noisy)      & & &&&1.000&0.006&0.007&0.009&0.009&0.037&0.007\\
(6): 3-3 (noisy)   & & &&&&1.000&0.016&0.048&0.065&0.313&0.618\\
    \addlinespace
(7): 1-1 (unnoisy)    & & &&&&&1.000&0.734&0.670&0.012&0.017\\
(8): 1-2 (unnoisy)       & & &&&&&&1.000&0.908&0.084&0.086\\
(9): 1-3 (unnoisy)   & & &&&&&&&1.000&0.125&0.132\\
(10): 1-1 (unnoisy)       & & &&&&&&&&1.000&0.587\\
(11): 1-2 (unnoisy)   & & &&&&&&&&&1.000\\
    \bottomrule
\end{tabular*}
\label{tab:author_rank_pos_0.9}
    \end{table}

    \begin{table}
    \caption{Node RBO similarity based on node strength for author networks created using positive tweets (p-value = 0.99)}
    \setlength\tabcolsep{0pt}
\begin{tabular*}{\linewidth}{@{\extracolsep{\fill}}
                                l
                           *{11}{c}
                            }
    \toprule
                & (1)   & (2)   & (3) & (4) & (5) & (6) & (7) & (8) & (9) & (10) & (11) \\
    \midrule
(1): 1-1 (noisy)    &1.000 &0.720&0.671&0.168&0.174&0.171&0.138&0.149&0.151&0.216&0.234\\
(2): 1-2 (noisy)     & &1.000 &0.865&0.233&0.244&0.189&0.170&0.207&0.207&0.281&0.281\\
(3): 1-3 (noisy)   & & &1.000&0.257&0.268&0.195&0.185&0.228&0.233&0.318&0.313\\
(4): 2-2  (noisy)  & & &&1.000&0.916&0.243&0.250&0.267&0.270&0.384&0.365\\
(5): 2-3  (noisy)      & & &&&1.000&0.265&0.253&0.277&0.283&0.401&0.384\\
(6): 3-3 (noisy)   & & &&&&1.000&0.251&0.297&0.316&0.395&0.470\\
    \addlinespace
(7): 1-1 (unnoisy)    & & &&&&&1.000&0.852&0.818&0.294&0.303\\
(8): 1-2 (unnoisy)       & & &&&&&&1.000&0.936&0.396&0.399\\
(9): 1-3 (unnoisy)   & & &&&&&&&1.000&0.421&0.431\\
(10): 1-1 (unnoisy)       & & &&&&&&&&1.000&0.865\\
(11): 1-2 (unnoisy)   & & &&&&&&&&&1.000\\
    \bottomrule
\end{tabular*}
\label{tab:author_rank_pos_0.99}
    \end{table}

\bigskip
\noindent
\textbf{Negative tweets}

Tables \ref{tab:author_rank_neg_0.9} and \ref{tab:author_rank_neg_0.99} show, respectively, the RBO similarities for the negative author networks node rankings for p-values of 0.9 and 0.99. As was the case in the co-occurrence networks, the similarities in the negative networks are different from those in the positive networks. In this case, however, the differences are now significantly more pronounced.

In general, the similarities observed in the negative tweets are higher than those in the equivalent positive tweet networks. This is not strictly true, but can be seen in the majority of comparisons. These differences can be quite high: for example, row (1) column (6) in Tables \ref{tab:author_rank_pos_0.9} and \ref{tab:author_rank_neg_0.9} are 0.009 and 0.236, and represent the equivalent comparisons in terms of data preprocessing setup. The trend of higher similarities for a higher p-value is once again seen in the negative tweet networks, but still with generally higher similarities overall.

This shows that it is not only the types of network that is constructed that can influence the sensitivity of rankings to data preprocessing, it is also the dataset itself. It may be that the topics discussed by negative tweets are more consistent and thus author similarities are similar. However, the differences between positive and negative tweet networks are far more clear in the author networks than the word co-occurrence networks, indicating that there is potentially some extra complexity in these results.

    \begin{table}
    \caption{Node RBO similarity based on node strength for author networks created using negative tweets (p-value = 0.9)}
    \setlength\tabcolsep{0pt}
\begin{tabular*}{\linewidth}{@{\extracolsep{\fill}}
                                l
                           *{11}{c}
                            }
    \toprule
                & (1)   & (2)   & (3) & (4) & (5) & (6) & (7) & (8) & (9) & (10) & (11) \\
    \midrule
(1): 1-1 (noisy)    &1.000 &0.744 &0.701&0.043&0.057&0.236&0.200&0.192&0.233&0.052&0.067\\
(2): 1-2 (noisy)     & &1.000 &0.873&0.107&0.116&0.229&0.165&0.168&0.217&0.113&0.121\\
(3): 1-3 (noisy)   & & &1.000&0.129&0.142&0.165&0.216&0.219&0.269&0.138&0.146\\
(4): 2-2  (noisy)  & & &&1.000&0.959&0.004&0.001&0.002&0.003&0.942&0.931\\
(5): 2-3  (noisy)      & & &&&1.000&0.015&0.001&0.002&0.002&0.953&0.966\\
(6): 3-3 (noisy)   & & &&&&1.000&0.002&0.003&0.001&0.012&0.023\\
    \addlinespace
(7): 1-1 (unnoisy)    & & &&&&&1.000&0.832&0.811&0.001&0.001\\
(8): 1-2 (unnoisy)       & & &&&&&&1.000&0.907&0.002&0.002\\
(9): 1-3 (unnoisy)   & & &&&&&&&1.000&0.002&0.002\\
(10): 1-1 (unnoisy)       & & &&&&&&&&1.000&0.969\\
(11): 1-2 (unnoisy)   & & &&&&&&&&&1.000\\
    \bottomrule
\end{tabular*}
\label{tab:author_rank_neg_0.9}
    \end{table}

    \begin{table}
    \caption{Node RBO similarity based on node strength for author networks created using negative tweets (p-value = 0.99)}
    \setlength\tabcolsep{0pt}
\begin{tabular*}{\linewidth}{@{\extracolsep{\fill}}
                                l
                           *{11}{c}
                            }
    \toprule
                & (1)   & (2)   & (3) & (4) & (5) & (6) & (7) & (8) & (9) & (10) & (11) \\
    \midrule
(1): 1-1 (noisy)    &1.000 &0.867&0.839&0.337&0.353&0.257&0.492&0.508&0.507&0.360&0.368\\
(2): 1-2 (noisy)     & &1.000 &0.930&0.385&0.399&0.256&0.483&0.509&0.513&0.405&0.410\\
(3): 1-3 (noisy)   & & &1.000&0.407&0.426&0.233&0.496&0.524&0.531&0.428&0.433\\
(4): 2-2  (noisy)  & & &&1.000&0.924&0.233&0.210&0.237&0.245&0.756&0.746\\
(5): 2-3  (noisy)      & & &&&1.000&0.268&0.215&0.243&0.250&0.780&0.790\\
(6): 3-3 (noisy)   & & &&&&1.000&0.176&0.179&0.170&0.265&0.300\\
    \addlinespace
(7): 1-1 (unnoisy)    & & &&&&&1.000&0.910&0.892&0.244&0.247\\
(8): 1-2 (unnoisy)       & & &&&&&&1.000&0.947&0.272&0.272\\
(9): 1-3 (unnoisy)   & & &&&&&&&1.000&0.277&0.278\\
(10): 1-1 (unnoisy)       & & &&&&&&&&1.000&0.933\\
(11): 1-2 (unnoisy)   & & &&&&&&&&&1.000\\
    \bottomrule
\end{tabular*}
\label{tab:author_rank_neg_0.99}
    \end{table}

\end{appendices}

\newpage 

\bibliography{sn-bibliography}


\begin{thebibliography}{47}
\ifx \bisbn   \undefined \def \bisbn  #1{ISBN #1}\fi
\ifx \binits  \undefined \def \binits#1{#1}\fi
\ifx \bauthor  \undefined \def \bauthor#1{#1}\fi
\ifx \batitle  \undefined \def \batitle#1{#1}\fi
\ifx \bjtitle  \undefined \def \bjtitle#1{#1}\fi
\ifx \bvolume  \undefined \def \bvolume#1{\textbf{#1}}\fi
\ifx \byear  \undefined \def \byear#1{#1}\fi
\ifx \bissue  \undefined \def \bissue#1{#1}\fi
\ifx \bfpage  \undefined \def \bfpage#1{#1}\fi
\ifx \blpage  \undefined \def \blpage #1{#1}\fi
\ifx \burl  \undefined \def \burl#1{\textsf{#1}}\fi
\ifx \doiurl  \undefined \def \doiurl#1{\url{https://doi.org/#1}}\fi
\ifx \betal  \undefined \def \betal{\textit{et al.}}\fi
\ifx \binstitute  \undefined \def \binstitute#1{#1}\fi
\ifx \binstitutionaled  \undefined \def \binstitutionaled#1{#1}\fi
\ifx \bctitle  \undefined \def \bctitle#1{#1}\fi
\ifx \beditor  \undefined \def \beditor#1{#1}\fi
\ifx \bpublisher  \undefined \def \bpublisher#1{#1}\fi
\ifx \bbtitle  \undefined \def \bbtitle#1{#1}\fi
\ifx \bedition  \undefined \def \bedition#1{#1}\fi
\ifx \bseriesno  \undefined \def \bseriesno#1{#1}\fi
\ifx \blocation  \undefined \def \blocation#1{#1}\fi
\ifx \bsertitle  \undefined \def \bsertitle#1{#1}\fi
\ifx \bsnm \undefined \def \bsnm#1{#1}\fi
\ifx \bsuffix \undefined \def \bsuffix#1{#1}\fi
\ifx \bparticle \undefined \def \bparticle#1{#1}\fi
\ifx \barticle \undefined \def \barticle#1{#1}\fi
\bibcommenthead
\ifx \bconfdate \undefined \def \bconfdate #1{#1}\fi
\ifx \botherref \undefined \def \botherref #1{#1}\fi
\ifx \url \undefined \def \url#1{\textsf{#1}}\fi
\ifx \bchapter \undefined \def \bchapter#1{#1}\fi
\ifx \bbook \undefined \def \bbook#1{#1}\fi
\ifx \bcomment \undefined \def \bcomment#1{#1}\fi
\ifx \oauthor \undefined \def \oauthor#1{#1}\fi
\ifx \citeauthoryear \undefined \def \citeauthoryear#1{#1}\fi
\ifx \endbibitem  \undefined \def \endbibitem {}\fi
\ifx \bconflocation  \undefined \def \bconflocation#1{#1}\fi
\ifx \arxivurl  \undefined \def \arxivurl#1{\textsf{#1}}\fi
\csname PreBibitemsHook\endcsname

\bibitem[\protect\citeauthoryear{Bono et~al.}{2023}]{bono2023pipeline}
\begin{barticle}
\bauthor{\bsnm{Bono}, \binits{C.A.}},
\bauthor{\bsnm{Cappiello}, \binits{C.}},
\bauthor{\bsnm{Pernici}, \binits{B.}},
\bauthor{\bsnm{Ramalli}, \binits{E.}},
\bauthor{\bsnm{Vitali}, \binits{M.}}:
\batitle{Pipeline design for data preparation for social media analysis}.
\bjtitle{ACM Journal of Data and Information Quality}
\bvolume{15}(\bissue{4}),
\bfpage{1}--\blpage{25}
(\byear{2023})
\end{barticle}
\endbibitem

\bibitem[\protect\citeauthoryear{Pastrav and Dignum}{2020}]{pastrav2020norms}
\begin{bchapter}
\bauthor{\bsnm{Pastrav}, \binits{C.}},
\bauthor{\bsnm{Dignum}, \binits{F.}}:
\bctitle{Norms in social simulation: balancing between realism and scalability}.
In: \bbtitle{Advances in Social Simulation: Looking in the Mirror},
pp. \bfpage{329}--\blpage{342}
(\byear{2020}).
\bcomment{Springer}
\end{bchapter}
\endbibitem

\bibitem[\protect\citeauthoryear{Fischer et~al.}{2005}]{fischer2005socionics}
\begin{bbook}
\bauthor{\bsnm{Fischer}, \binits{K.}},
\bauthor{\bsnm{Florian}, \binits{M.}},
\bauthor{\bsnm{Malsch}, \binits{T.}}:
\bbtitle{Socionics: Scalability of Complex Social Systems}
vol. \bseriesno{3413}.
\bpublisher{Springer}, \blocation{???}
(\byear{2005})
\end{bbook}
\endbibitem

\bibitem[\protect\citeauthoryear{Kim et~al.}{2016}]{kim2016examples}
\begin{botherref}
\oauthor{\bsnm{Kim}, \binits{B.}},
\oauthor{\bsnm{Khanna}, \binits{R.}},
\oauthor{\bsnm{Koyejo}, \binits{O.O.}}:
Examples are not enough, learn to criticize! criticism for interpretability.
Advances in neural information processing systems
\textbf{29}
(2016)
\end{botherref}
\endbibitem

\bibitem[\protect\citeauthoryear{Varshney and Vishwakarma}{2021}]{varshney2021review}
\begin{barticle}
\bauthor{\bsnm{Varshney}, \binits{D.}},
\bauthor{\bsnm{Vishwakarma}, \binits{D.K.}}:
\batitle{A review on rumour prediction and veracity assessment in online social network}.
\bjtitle{Expert Systems with Applications}
\bvolume{168},
\bfpage{114208}
(\byear{2021})
\end{barticle}
\endbibitem

\bibitem[\protect\citeauthoryear{Gunti et~al.}{2022}]{gunti2022data}
\begin{bchapter}
\bauthor{\bsnm{Gunti}, \binits{P.}},
\bauthor{\bsnm{Gupta}, \binits{B.B.}},
\bauthor{\bsnm{Benkhelifa}, \binits{E.}}:
\bctitle{Data mining approaches for sentiment analysis in online social networks (osns)}.
In: \bbtitle{Data Mining Approaches for Big Data and Sentiment Analysis in Social Media},
pp. \bfpage{116}--\blpage{141}.
\bpublisher{IGI Global}, \blocation{???}
(\byear{2022})
\end{bchapter}
\endbibitem

\bibitem[\protect\citeauthoryear{Kosinski et~al.}{2014}]{kosinski2014manifestations}
\begin{barticle}
\bauthor{\bsnm{Kosinski}, \binits{M.}},
\bauthor{\bsnm{Bachrach}, \binits{Y.}},
\bauthor{\bsnm{Kohli}, \binits{P.}},
\bauthor{\bsnm{Stillwell}, \binits{D.}},
\bauthor{\bsnm{Graepel}, \binits{T.}}:
\batitle{Manifestations of user personality in website choice and behaviour on online social networks}.
\bjtitle{Machine learning}
\bvolume{95},
\bfpage{357}--\blpage{380}
(\byear{2014})
\end{barticle}
\endbibitem

\bibitem[\protect\citeauthoryear{Pourmand et~al.}{2019}]{pourmand2019social}
\begin{barticle}
\bauthor{\bsnm{Pourmand}, \binits{A.}},
\bauthor{\bsnm{Roberson}, \binits{J.}},
\bauthor{\bsnm{Caggiula}, \binits{A.}},
\bauthor{\bsnm{Monsalve}, \binits{N.}},
\bauthor{\bsnm{Rahimi}, \binits{M.}},
\bauthor{\bsnm{Torres-Llenza}, \binits{V.}}:
\batitle{Social media and suicide: a review of technology-based epidemiology and risk assessment}.
\bjtitle{Telemedicine and e-Health}
\bvolume{25}(\bissue{10}),
\bfpage{880}--\blpage{888}
(\byear{2019})
\end{barticle}
\endbibitem

\bibitem[\protect\citeauthoryear{Prasad et~al.}{2019}]{prasad2019purchase}
\begin{barticle}
\bauthor{\bsnm{Prasad}, \binits{S.}},
\bauthor{\bsnm{Garg}, \binits{A.}},
\bauthor{\bsnm{Prasad}, \binits{S.}}:
\batitle{Purchase decision of generation y in an online environment}.
\bjtitle{Marketing Intelligence \& Planning}
\bvolume{37}(\bissue{4}),
\bfpage{372}--\blpage{385}
(\byear{2019})
\end{barticle}
\endbibitem

\bibitem[\protect\citeauthoryear{Cohen and Havlin}{2010}]{cohen2010complex}
\begin{bbook}
\bauthor{\bsnm{Cohen}, \binits{R.}},
\bauthor{\bsnm{Havlin}, \binits{S.}}:
\bbtitle{Complex Networks: Structure, Robustness and Function}.
\bpublisher{Cambridge university press}, \blocation{???}
(\byear{2010})
\end{bbook}
\endbibitem

\bibitem[\protect\citeauthoryear{Garg and Kumar}{2018}]{garg2018structure}
\begin{barticle}
\bauthor{\bsnm{Garg}, \binits{M.}},
\bauthor{\bsnm{Kumar}, \binits{M.}}:
\batitle{The structure of word co-occurrence network for microblogs}.
\bjtitle{Physica A: Statistical Mechanics and its Applications}
\bvolume{512},
\bfpage{698}--\blpage{720}
(\byear{2018})
\end{barticle}
\endbibitem

\bibitem[\protect\citeauthoryear{Fudolig et~al.}{2022}]{fudolig2022sentiment}
\begin{barticle}
\bauthor{\bsnm{Fudolig}, \binits{M.I.}},
\bauthor{\bsnm{Alshaabi}, \binits{T.}},
\bauthor{\bsnm{Arnold}, \binits{M.V.}},
\bauthor{\bsnm{Danforth}, \binits{C.M.}},
\bauthor{\bsnm{Dodds}, \binits{P.S.}}:
\batitle{Sentiment and structure in word co-occurrence networks on twitter}.
\bjtitle{Applied Network Science}
\bvolume{7}(\bissue{1}),
\bfpage{1}--\blpage{27}
(\byear{2022})
\end{barticle}
\endbibitem

\bibitem[\protect\citeauthoryear{Lozano et~al.}{2019}]{lozano2019complex}
\begin{barticle}
\bauthor{\bsnm{Lozano}, \binits{S.}},
\bauthor{\bsnm{Calzada-Infante}, \binits{L.}},
\bauthor{\bsnm{Adenso-D{\'\i}az}, \binits{B.}},
\bauthor{\bsnm{Garc{\'\i}a}, \binits{S.}}:
\batitle{Complex network analysis of keywords co-occurrence in the recent efficiency analysis literature}.
\bjtitle{Scientometrics}
\bvolume{120},
\bfpage{609}--\blpage{629}
(\byear{2019})
\end{barticle}
\endbibitem

\bibitem[\protect\citeauthoryear{Nevin et~al.}{2021}]{nevin2021non}
\begin{botherref}
\oauthor{\bsnm{Nevin}, \binits{J.}},
\oauthor{\bsnm{Lees}, \binits{M.}},
\oauthor{\bsnm{Groth}, \binits{P.}}:
The non-linear impact of data handling on network diffusion models.
Patterns
\textbf{2}(12)
(2021)
\end{botherref}
\endbibitem

\bibitem[\protect\citeauthoryear{Nevin et~al.}{2023a}]{nevin2023approach}
\begin{barticle}
\bauthor{\bsnm{Nevin}, \binits{J.}},
\bauthor{\bsnm{Groth}, \binits{P.}},
\bauthor{\bsnm{Lees}, \binits{M.}}:
\batitle{An approach for analysing the impact of data integration on complex network diffusion models}.
\bjtitle{Journal of Complex Networks}
\bvolume{11}(\bissue{4}),
\bfpage{025}
(\byear{2023})
\end{barticle}
\endbibitem

\bibitem[\protect\citeauthoryear{Nevin et~al.}{2023b}]{nevin2023data}
\begin{bchapter}
\bauthor{\bsnm{Nevin}, \binits{J.}},
\bauthor{\bsnm{Groth}, \binits{P.}},
\bauthor{\bsnm{Lees}, \binits{M.}}:
\bctitle{Data integration landscapes: The case for non-optimal solutions in network diffusion models}.
In: \bbtitle{International Conference on Computational Science},
pp. \bfpage{494}--\blpage{508}
(\byear{2023}).
\bcomment{Springer}
\end{bchapter}
\endbibitem

\bibitem[\protect\citeauthoryear{Dorogovtsev and Mendes}{2022}]{dorogovtsev2022nature}
\begin{bbook}
\bauthor{\bsnm{Dorogovtsev}, \binits{S.N.}},
\bauthor{\bsnm{Mendes}, \binits{J.F.}}:
\bbtitle{The Nature of Complex Networks}.
\bpublisher{Oxford University Press}, \blocation{???}
(\byear{2022})
\end{bbook}
\endbibitem

\bibitem[\protect\citeauthoryear{Boschetti et~al.}{2005}]{boschetti2005defining}
\begin{bchapter}
\bauthor{\bsnm{Boschetti}, \binits{F.}},
\bauthor{\bsnm{Prokopenko}, \binits{M.}},
\bauthor{\bsnm{Macreadie}, \binits{I.}},
\bauthor{\bsnm{Grisogono}, \binits{A.-M.}}:
\bctitle{Defining and detecting emergence in complex networks}.
In: \bbtitle{International Conference on Knowledge-Based and Intelligent Information and Engineering Systems},
pp. \bfpage{573}--\blpage{580}
(\byear{2005}).
\bcomment{Springer}
\end{bchapter}
\endbibitem

\bibitem[\protect\citeauthoryear{Delgado}{2002}]{delgado2002emergence}
\begin{barticle}
\bauthor{\bsnm{Delgado}, \binits{J.}}:
\batitle{Emergence of social conventions in complex networks}.
\bjtitle{Artificial intelligence}
\bvolume{141}(\bissue{1-2}),
\bfpage{171}--\blpage{185}
(\byear{2002})
\end{barticle}
\endbibitem

\bibitem[\protect\citeauthoryear{Mata}{2020}]{mata2020complex}
\begin{barticle}
\bauthor{\bsnm{Mata}, \binits{A.S.d.}}:
\batitle{Complex networks: a mini-review}.
\bjtitle{Brazilian Journal of Physics}
\bvolume{50},
\bfpage{658}--\blpage{672}
(\byear{2020})
\end{barticle}
\endbibitem

\bibitem[\protect\citeauthoryear{Barab{\'a}si}{2013}]{barabasi2013network}
\begin{barticle}
\bauthor{\bsnm{Barab{\'a}si}, \binits{A.-L.}}:
\batitle{Network science}.
\bjtitle{Philosophical Transactions of the Royal Society A: Mathematical, Physical and Engineering Sciences}
\bvolume{371}(\bissue{1987}),
\bfpage{20120375}
(\byear{2013})
\end{barticle}
\endbibitem

\bibitem[\protect\citeauthoryear{Staudt et~al.}{2016}]{staudt2016networkit}
\begin{barticle}
\bauthor{\bsnm{Staudt}, \binits{C.L.}},
\bauthor{\bsnm{Sazonovs}, \binits{A.}},
\bauthor{\bsnm{Meyerhenke}, \binits{H.}}:
\batitle{Networkit: A tool suite for large-scale complex network analysis}.
\bjtitle{Network Science}
\bvolume{4}(\bissue{4}),
\bfpage{508}--\blpage{530}
(\byear{2016})
\end{barticle}
\endbibitem

\bibitem[\protect\citeauthoryear{Barab{\'a}si and Bonabeau}{2003}]{barabasi2003scale}
\begin{barticle}
\bauthor{\bsnm{Barab{\'a}si}, \binits{A.-L.}},
\bauthor{\bsnm{Bonabeau}, \binits{E.}}:
\batitle{Scale-free networks}.
\bjtitle{Scientific american}
\bvolume{288}(\bissue{5}),
\bfpage{60}--\blpage{69}
(\byear{2003})
\end{barticle}
\endbibitem

\bibitem[\protect\citeauthoryear{Amaral et~al.}{2000}]{amaral2000classes}
\begin{barticle}
\bauthor{\bsnm{Amaral}, \binits{L.A.N.}},
\bauthor{\bsnm{Scala}, \binits{A.}},
\bauthor{\bsnm{Barthelemy}, \binits{M.}},
\bauthor{\bsnm{Stanley}, \binits{H.E.}}:
\batitle{Classes of small-world networks}.
\bjtitle{Proceedings of the national academy of sciences}
\bvolume{97}(\bissue{21}),
\bfpage{11149}--\blpage{11152}
(\byear{2000})
\end{barticle}
\endbibitem

\bibitem[\protect\citeauthoryear{Sciarra et~al.}{2018}]{sciarra2018change}
\begin{barticle}
\bauthor{\bsnm{Sciarra}, \binits{C.}},
\bauthor{\bsnm{Chiarotti}, \binits{G.}},
\bauthor{\bsnm{Laio}, \binits{F.}},
\bauthor{\bsnm{Ridolfi}, \binits{L.}}:
\batitle{A change of perspective in network centrality}.
\bjtitle{Scientific reports}
\bvolume{8}(\bissue{1}),
\bfpage{15269}
(\byear{2018})
\end{barticle}
\endbibitem

\bibitem[\protect\citeauthoryear{Salavati et~al.}{2019}]{salavati2019ranking}
\begin{barticle}
\bauthor{\bsnm{Salavati}, \binits{C.}},
\bauthor{\bsnm{Abdollahpouri}, \binits{A.}},
\bauthor{\bsnm{Manbari}, \binits{Z.}}:
\batitle{Ranking nodes in complex networks based on local structure and improving closeness centrality}.
\bjtitle{Neurocomputing}
\bvolume{336},
\bfpage{36}--\blpage{45}
(\byear{2019})
\end{barticle}
\endbibitem

\bibitem[\protect\citeauthoryear{Wang et~al.}{2022}]{wang2022identifying}
\begin{barticle}
\bauthor{\bsnm{Wang}, \binits{Y.}},
\bauthor{\bsnm{Li}, \binits{H.}},
\bauthor{\bsnm{Zhang}, \binits{L.}},
\bauthor{\bsnm{Zhao}, \binits{L.}},
\bauthor{\bsnm{Li}, \binits{W.}}:
\batitle{Identifying influential nodes in social networks: Centripetal centrality and seed exclusion approach}.
\bjtitle{Chaos, Solitons \& Fractals}
\bvolume{162},
\bfpage{112513}
(\byear{2022})
\end{barticle}
\endbibitem

\bibitem[\protect\citeauthoryear{Ugurlu}{2022}]{ugurlu2022comparative}
\begin{barticle}
\bauthor{\bsnm{Ugurlu}, \binits{O.}}:
\batitle{Comparative analysis of centrality measures for identifying critical nodes in complex networks}.
\bjtitle{Journal of Computational Science}
\bvolume{62},
\bfpage{101738}
(\byear{2022})
\end{barticle}
\endbibitem

\bibitem[\protect\citeauthoryear{Malik}{2022}]{malik2022complex}
\begin{barticle}
\bauthor{\bsnm{Malik}, \binits{H.A.M.}}:
\batitle{Complex network formation and analysis of online social media systems}.
\bjtitle{Cmes-Comr Model Engg \& Sci}
\bvolume{130}(\bissue{3}),
\bfpage{1737}--\blpage{1750}
(\byear{2022})
\end{barticle}
\endbibitem

\bibitem[\protect\citeauthoryear{Mocibob et~al.}{2016}]{mocibob2016revealing}
\begin{bchapter}
\bauthor{\bsnm{Mocibob}, \binits{E.}},
\bauthor{\bsnm{Martin{\v{c}}i{\'c}-Ip{\v{s}}i{\'c}}, \binits{S.}},
\bauthor{\bsnm{Me{\v{s}}trovi{\'c}}, \binits{A.}}:
\bctitle{Revealing the structure of domain specific tweets via complex networks analysis}.
In: \bbtitle{2016 39th International Convention on Information and Communication Technology, Electronics and Microelectronics (MIPRO)},
pp. \bfpage{1623}--\blpage{1627}
(\byear{2016}).
\bcomment{IEEE}
\end{bchapter}
\endbibitem

\bibitem[\protect\citeauthoryear{Moreno-Ortiz and Garc{\'\i}a-G{\'a}mez}{2023}]{moreno2023strategies}
\begin{botherref}
\oauthor{\bsnm{Moreno-Ortiz}, \binits{A.}},
\oauthor{\bsnm{Garc{\'\i}a-G{\'a}mez}, \binits{M.}}:
Strategies for the analysis of large social media corpora: Sampling and keyword extraction methods.
Corpus Pragmatics,
1--25
(2023)
\end{botherref}
\endbibitem

\bibitem[\protect\citeauthoryear{Jayasiriwardene and Ganegoda}{2020}]{jayasiriwardene2020keyword}
\begin{bchapter}
\bauthor{\bsnm{Jayasiriwardene}, \binits{T.D.}},
\bauthor{\bsnm{Ganegoda}, \binits{G.U.}}:
\bctitle{Keyword extraction from tweets using nlp tools for collecting relevant news}.
In: \bbtitle{2020 International Research Conference on Smart Computing and Systems Engineering (SCSE)},
pp. \bfpage{129}--\blpage{135}
(\byear{2020}).
\bcomment{IEEE}
\end{bchapter}
\endbibitem

\bibitem[\protect\citeauthoryear{Li et~al.}{2016}]{li2016evolutionary}
\begin{barticle}
\bauthor{\bsnm{Li}, \binits{H.}},
\bauthor{\bsnm{An}, \binits{H.}},
\bauthor{\bsnm{Wang}, \binits{Y.}},
\bauthor{\bsnm{Huang}, \binits{J.}},
\bauthor{\bsnm{Gao}, \binits{X.}}:
\batitle{Evolutionary features of academic articles co-keyword network and keywords co-occurrence network: Based on two-mode affiliation network}.
\bjtitle{Physica A: Statistical Mechanics and its Applications}
\bvolume{450},
\bfpage{657}--\blpage{669}
(\byear{2016})
\end{barticle}
\endbibitem

\bibitem[\protect\citeauthoryear{Radhakrishnan et~al.}{2017}]{radhakrishnan2017novel}
\begin{barticle}
\bauthor{\bsnm{Radhakrishnan}, \binits{S.}},
\bauthor{\bsnm{Erbis}, \binits{S.}},
\bauthor{\bsnm{Isaacs}, \binits{J.A.}},
\bauthor{\bsnm{Kamarthi}, \binits{S.}}:
\batitle{Novel keyword co-occurrence network-based methods to foster systematic reviews of scientific literature}.
\bjtitle{PloS one}
\bvolume{12}(\bissue{3}),
\bfpage{0172778}
(\byear{2017})
\end{barticle}
\endbibitem

\bibitem[\protect\citeauthoryear{Fu and Waltman}{2022}]{fu2022large}
\begin{barticle}
\bauthor{\bsnm{Fu}, \binits{H.-Z.}},
\bauthor{\bsnm{Waltman}, \binits{L.}}:
\batitle{A large-scale bibliometric analysis of global climate change research between 2001 and 2018}.
\bjtitle{Climatic Change}
\bvolume{170}(\bissue{3-4}),
\bfpage{36}
(\byear{2022})
\end{barticle}
\endbibitem

\bibitem[\protect\citeauthoryear{Van~Vliet et~al.}{2020}]{van2020twitter}
\begin{barticle}
\bauthor{\bsnm{Van~Vliet}, \binits{L.}},
\bauthor{\bsnm{T{\"o}rnberg}, \binits{P.}},
\bauthor{\bsnm{Uitermark}, \binits{J.}}:
\batitle{The twitter parliamentarian database: Analyzing twitter politics across 26 countries}.
\bjtitle{PLoS one}
\bvolume{15}(\bissue{9}),
\bfpage{0237073}
(\byear{2020})
\end{barticle}
\endbibitem

\bibitem[\protect\citeauthoryear{Falkenberg et~al.}{2022}]{falkenberg2022growing}
\begin{barticle}
\bauthor{\bsnm{Falkenberg}, \binits{M.}},
\bauthor{\bsnm{Galeazzi}, \binits{A.}},
\bauthor{\bsnm{Torricelli}, \binits{M.}},
\bauthor{\bsnm{Di~Marco}, \binits{N.}},
\bauthor{\bsnm{Larosa}, \binits{F.}},
\bauthor{\bsnm{Sas}, \binits{M.}},
\bauthor{\bsnm{Mekacher}, \binits{A.}},
\bauthor{\bsnm{Pearce}, \binits{W.}},
\bauthor{\bsnm{Zollo}, \binits{F.}},
\bauthor{\bsnm{Quattrociocchi}, \binits{W.}}, \betal:
\batitle{Growing polarization around climate change on social media}.
\bjtitle{Nature Climate Change}
\bvolume{12}(\bissue{12}),
\bfpage{1114}--\blpage{1121}
(\byear{2022})
\end{barticle}
\endbibitem

\bibitem[\protect\citeauthoryear{Esteve Del~Valle et~al.}{2022}]{esteve2022political}
\begin{barticle}
\bauthor{\bsnm{Esteve Del~Valle}, \binits{M.}},
\bauthor{\bsnm{Broersma}, \binits{M.}},
\bauthor{\bsnm{Ponsioen}, \binits{A.}}:
\batitle{Political interaction beyond party lines: Communication ties and party polarization in parliamentary twitter networks}.
\bjtitle{Social science computer review}
\bvolume{40}(\bissue{3}),
\bfpage{736}--\blpage{755}
(\byear{2022})
\end{barticle}
\endbibitem

\bibitem[\protect\citeauthoryear{Praet et~al.}{2021}]{praet2021patterns}
\begin{barticle}
\bauthor{\bsnm{Praet}, \binits{S.}},
\bauthor{\bsnm{Martens}, \binits{D.}},
\bauthor{\bsnm{Van~Aelst}, \binits{P.}}:
\batitle{Patterns of democracy? social network analysis of parliamentary twitter networks in 12 countries}.
\bjtitle{Online Social Networks and Media}
\bvolume{24},
\bfpage{100154}
(\byear{2021})
\end{barticle}
\endbibitem

\bibitem[\protect\citeauthoryear{Torres-Salinas et~al.}{2009}]{torres2009ranking}
\begin{barticle}
\bauthor{\bsnm{Torres-Salinas}, \binits{D.}},
\bauthor{\bsnm{Lopez-C{\'o}zar}, \binits{E.}},
\bauthor{\bsnm{Jim{\'e}nez-Contreras}, \binits{E.}}:
\batitle{Ranking of departments and researchers within a university using two different databases: Web of science versus scopus}.
\bjtitle{Scientometrics}
\bvolume{80}(\bissue{3}),
\bfpage{761}--\blpage{774}
(\byear{2009})
\end{barticle}
\endbibitem

\bibitem[\protect\citeauthoryear{Archambault et~al.}{2009}]{archambault2009comparing}
\begin{barticle}
\bauthor{\bsnm{Archambault}, \binits{{\'E}.}},
\bauthor{\bsnm{Campbell}, \binits{D.}},
\bauthor{\bsnm{Gingras}, \binits{Y.}},
\bauthor{\bsnm{Larivi{\`e}re}, \binits{V.}}:
\batitle{Comparing bibliometric statistics obtained from the web of science and scopus}.
\bjtitle{Journal of the American society for information science and technology}
\bvolume{60}(\bissue{7}),
\bfpage{1320}--\blpage{1326}
(\byear{2009})
\end{barticle}
\endbibitem

\bibitem[\protect\citeauthoryear{Rose et~al.}{2010}]{rose2010automatic}
\begin{botherref}
\oauthor{\bsnm{Rose}, \binits{S.}},
\oauthor{\bsnm{Engel}, \binits{D.}},
\oauthor{\bsnm{Cramer}, \binits{N.}},
\oauthor{\bsnm{Cowley}, \binits{W.}}:
Automatic keyword extraction from individual documents.
Text mining: applications and theory,
1--20
(2010)
\end{botherref}
\endbibitem

\bibitem[\protect\citeauthoryear{Hutto and Gilbert}{2014}]{hutto2014vader}
\begin{bchapter}
\bauthor{\bsnm{Hutto}, \binits{C.}},
\bauthor{\bsnm{Gilbert}, \binits{E.}}:
\bctitle{Vader: A parsimonious rule-based model for sentiment analysis of social media text}.
In: \bbtitle{Proceedings of the International AAAI Conference on Web and Social Media},
vol. \bseriesno{8},
pp. \bfpage{216}--\blpage{225}
(\byear{2014})
\end{bchapter}
\endbibitem

\bibitem[\protect\citeauthoryear{Bird et~al.}{2009}]{bird2009natural}
\begin{bbook}
\bauthor{\bsnm{Bird}, \binits{S.}},
\bauthor{\bsnm{Klein}, \binits{E.}},
\bauthor{\bsnm{Loper}, \binits{E.}}:
\bbtitle{Natural Language Processing with Python: Analyzing Text with the Natural Language Toolkit}.
\bpublisher{" O'Reilly Media, Inc."}, \blocation{???}
(\byear{2009})
\end{bbook}
\endbibitem

\bibitem[\protect\citeauthoryear{Chen et~al.}{2012}]{chen2012identifying}
\begin{barticle}
\bauthor{\bsnm{Chen}, \binits{D.}},
\bauthor{\bsnm{L{\"u}}, \binits{L.}},
\bauthor{\bsnm{Shang}, \binits{M.-S.}},
\bauthor{\bsnm{Zhang}, \binits{Y.-C.}},
\bauthor{\bsnm{Zhou}, \binits{T.}}:
\batitle{Identifying influential nodes in complex networks}.
\bjtitle{Physica a: Statistical mechanics and its applications}
\bvolume{391}(\bissue{4}),
\bfpage{1777}--\blpage{1787}
(\byear{2012})
\end{barticle}
\endbibitem

\bibitem[\protect\citeauthoryear{L{\"u} et~al.}{2016}]{lu2016vital}
\begin{barticle}
\bauthor{\bsnm{L{\"u}}, \binits{L.}},
\bauthor{\bsnm{Chen}, \binits{D.}},
\bauthor{\bsnm{Ren}, \binits{X.-L.}},
\bauthor{\bsnm{Zhang}, \binits{Q.-M.}},
\bauthor{\bsnm{Zhang}, \binits{Y.-C.}},
\bauthor{\bsnm{Zhou}, \binits{T.}}:
\batitle{Vital nodes identification in complex networks}.
\bjtitle{Physics reports}
\bvolume{650},
\bfpage{1}--\blpage{63}
(\byear{2016})
\end{barticle}
\endbibitem

\bibitem[\protect\citeauthoryear{Webber et~al.}{2010}]{webber2010similarity}
\begin{barticle}
\bauthor{\bsnm{Webber}, \binits{W.}},
\bauthor{\bsnm{Moffat}, \binits{A.}},
\bauthor{\bsnm{Zobel}, \binits{J.}}:
\batitle{A similarity measure for indefinite rankings}.
\bjtitle{ACM Transactions on Information Systems (TOIS)}
\bvolume{28}(\bissue{4}),
\bfpage{1}--\blpage{38}
(\byear{2010})
\end{barticle}
\endbibitem

\end{thebibliography}

\end{document}